\documentclass[fleqn,usenatbib]{rasti}

\usepackage{newtxtext,newtxmath}
\usepackage[T1]{fontenc}
\usepackage{graphicx}%
\usepackage{subcaption}
\usepackage{caption}
\graphicspath{{figs/}}
\usepackage{bm,csquotes}
\usepackage{booktabs}
\usepackage{float}
\usepackage{tabularx}
\usepackage{siunitx}
\usepackage{geometry} % To adjust margins for a wide table
\usepackage{xcolor}   % Required for coloring rows
\usepackage{colortbl}
\usepackage{algorithm}%
\usepackage{algorithmicx}%
\usepackage{algpseudocode}%

\DeclareRobustCommand{\VAN}[3]{#2}
\let\VANthebibliography\thebibliography
\def\thebibliography{\DeclareRobustCommand{\VAN}[3]{##3}\VANthebibliography}

\usepackage{graphicx}	% Including figure files
\usepackage{amsmath}	% Advanced maths commands
\usepackage{changepage}

\title[Polarisation Primary Beam of GRAO 32-m Telescope]{Full Characterisation of the Polarisation Primary Beam of the GRAO 32-m Telescope}

\author[ T. Ansah-Narh et al.]{
Theophilus  Ansah-Narh,$^{1}$\thanks{E-mail: \href{theophilus.ansah-narh@gaec.gov.gh}{theophilus.ansah-narh@gaec.gov.gh} (TA-N)}
Nia Imara,$^{2}$
Benedicta Woode$^{1}$
Joseph Bremang Tandoh,$^{1}$
\newauthor
Emmanuel Proven Adzri,$^{1}$
Diana Klutse,$^{1}$
and Evaristus Uzochukwu Iyida$^{1, 3}$
\\
$^{1}$Ghana Space Science and Technology Institute, Ghana Atomic Energy Commission, P.O. Box LG 80, Legon-Accra, Ghana\\
$^{2}$Astronomy and Astrophysics Department, University of California, Santa Cruz, CA, USA\\
$^{3}$Department of Physics and Astronomy, Faculty of Physical Sciences, University of Nigeria, Nsukka, Nigeria.
}

\date{Accepted XXX. Received YYY; in original form ZZZ}

\pubyear{\the\year{}}

\begin{document}
\label{firstpage}
\pagerange{\pageref{firstpage}--\pageref{lastpage}}
\maketitle

\begin{abstract}
Direction-dependent instrumental polarisation is a major systematic
limitation in high-fidelity single-dish radio polarimetry, yet a unified
characterisation of beam leakage and the conditioning of polarisation
recovery is lacking for the Ghana Radio Astronomy Observatory (GRAO)
32-m telescope. We aim to establish a quantitative, direction-dependent
polarimetric beam model at 5.0 and 6.7~GHz. High-resolution
\texttt{GRASP} electromagnetic simulations of the nominal telescope
configuration are used to derive the full complex Jones response and
corresponding Mueller matrices, from which beam shape, beam squint,
instrumental Stokes leakage, and intrinsic cross-polarisation ratio
(IXR) are evaluated across the primary beam. The Stokes~$I$ beams have
half-power beam widths of $396.4$ and $295.3$~arcsec at 5.0 and
6.7~GHz, respectively, with main-beam efficiencies of $56.4$ and
$55.6$~per~cent. At 5.0~GHz, the circular-polarisation beam squint is
$9.96$~arcsec ($2.5$~per~cent of the HPBW), whereas no statistically
significant squint is detected at 6.7~GHz. Leakage from Stokes~$I$ into
linear polarisation remains below $0.1$~per~cent within the half-power
beam but increases substantially towards the sidelobes. IXR reaches
approximately $80$~dB on axis and decreases with angular offset.
These results establish the intrinsic electromagnetic polarimetric
response of the GRAO 32-m telescope and provide a quantitative baseline
for direction-dependent calibration and subsequent observational
validation.
\end{abstract}

% Include between one and six keywords.
\begin{keywords}
Instrumentation -- Software -- Primary beam modelling -- Ghana Radio Astronomy Observatory -- Radio polarimetry
\end{keywords}

\section{Introduction}
\label{sec:intro}

Radio polarimetry provides a uniquely powerful window into the
magnetised Universe, enabling direct probes of magnetic field structure
and strength across a wide range of astrophysical environments.
Linearly polarised synchrotron emission traces ordered magnetic fields
in the interstellar and intergalactic media, while Faraday rotation
encodes information on the line-of-sight magnetic field component and
thermal electron density. Together, these observables underpin
contemporary studies of Galactic magnetism, the evolution of magnetic
fields in galaxies and galaxy clusters, and the physical conditions in
active galactic nuclei and their relativistic jets \citep{Beck2015}. At
smaller spatial scales, polarimetric observations of maser emission
provide a direct diagnostic of magnetic fields in star-forming regions,
offering insight into the role of magnetism in the earliest stages of
stellar evolution. Beyond astrophysical magnetism, accurate
characterisation of polarised emission is also essential for cosmology,
where Galactic foreground polarisation constitutes a major systematic
in measurements of the cosmic microwave background and large-scale
structure \citep{aghanim2020planck}.

As instrumental sensitivities have steadily improved, the dominant
limitation in many polarimetric experiments has shifted from thermal
noise to systematic effects associated with the telescope response
itself. In particular, instrumental polarisation introduced by the
antenna and receiver system can produce spurious Stokes signals that
mimic or obscure weak astrophysical polarisation. This is especially
critical for single-dish telescopes, where the measured polarisation is
convolved with a direction-dependent beam response that varies across
the field of view. Unlike total-intensity measurements, which are often
adequately described by a scalar beam pattern, polarimetric observations
are governed by a full matrix response that couples the incident Stokes
parameters in a non-trivial manner.

Instrumental polarisation in single-dish systems arises from a
combination of optical and electromagnetic effects, including feed
imperfections, reflector geometry, support structures, and departures
from ideal symmetry. These effects manifest observationally as beam
squint, differential beam shapes between nominally orthogonal
polarisations, and cross-polarisation that varies with angular offset
from the boresight. As a result, the telescope response cannot be
represented by a single leakage term or constant calibration factor.
Instead, the polarisation response must be treated as a spatially
varying operator that mixes Stokes $I$, $Q$, $U$, and $V$ in a
direction-dependent way. Failure to account for this behaviour leads to
systematic corruption of the recovered Stokes parameters, particularly
when mapping extended sources or when high dynamic range polarimetry is
required.

Over the past two decades, significant progress has been made in
modelling and calibrating instrumental polarisation through the use of
Jones and Mueller matrix formalisms, which provide a rigorous framework
for describing the polarimetric response of radio telescopes
\citep{hamaker1996understanding}. For many instruments, beam models have
been used to quantify cross-polarisation levels or to report peak
instrumental leakage within the main beam. While such metrics are
informative, they provide only a partial description of polarimetric
performance. In particular, most existing beam characterisations do not
address the numerical stability of the polarisation inversion problem
itself, that is, how reliably the true sky polarisation can be recovered
from the measured voltages in the presence of an imperfect and
direction-dependent Jones matrix. This aspect is fundamentally linked
to the conditioning of the instrumental response and cannot be inferred
from leakage percentages alone.

Beam characterisation studies of other radio telescopes demonstrate the
importance of treating instrumental polarisation as a
direction-dependent property of the antenna response. For the Dominion
Radio Astrophysical Observatory (DRAO) Synthesis Telescope,
electromagnetic calculations and measurements have been used to
characterise instrumental polarisation in terms of $Q/I$, $U/I$, and
$V/I$ across the main beam and near sidelobes. These studies showed that
cross-polarisation associated with the feed and scattering from
structural elements can generate substantial instrumental polarisation
away from boresight, with particularly strong effects in the sidelobes
\citep{ng2005polarization}. The Arecibo telescope was similarly
characterised through an all-Stokes description of the main beam and
first sidelobe, including beamwidth, beam efficiency, beam squint, and
beam squash, demonstrating that polarimetric beam properties can differ
systematically from the total-intensity response \citep{Heiles2001}.
More recently, full-polarisation primary-beam measurements of MeerKAT
antennas using radio holography have quantified frequency-, pointing-,
and antenna-dependent variations across the half-power region,
illustrating the need for accurate beam models when direction-dependent
polarisation effects become important \citep{deVilliers2022}.
These studies establish that direction-dependent
instrumental polarisation is a general property of practical radio
telescopes and that accurate beam characterisation is essential for
high-fidelity polarimetric observations.

Despite these advances, published beam characterisations do not
generally employ a common set of metrics that simultaneously describes
the electromagnetic field response, Stokes leakage, beam displacement,
and the numerical conditioning of polarisation recovery. Most
observational and engineering studies emphasise beam shape,
cross-polarisation, instrumental leakage, beam squint, or related
quantities, while a formal assessment of the stability of the
polarisation inversion across the full primary beam is less commonly
presented. This distinction is important because two systems with
similar leakage levels can nevertheless exhibit different sensitivities
to noise and calibration errors if their underlying Jones matrices have
different condition numbers. A unified treatment that connects the
direction-dependent Jones response to Mueller leakage and polarimetric
conditioning therefore provides a more complete description of
telescope performance.

The intrinsic cross-polarisation ratio (IXR) provides a compact and
physically meaningful measure of this conditioning, expressing the
sensitivity of the polarimetric inversion to noise and calibration
errors in terms of the condition number of the Jones matrix
\citep{carozzi2011fundamental}. Unlike traditional leakage metrics, IXR
directly quantifies the stability of Stokes parameter recovery, thereby
providing a fundamental measure of the intrinsic polarimetric fidelity
of the instrumental response. IXR has been applied to established
radio-polarimetric systems, including measurements of receptor
non-orthogonality in instruments such as the Parkes telescope, where
values of approximately $40$--$60$~dB have been reported across the
observing band \citep{Hobbs2020}. However, IXR is less commonly
incorporated into direction-dependent primary-beam characterisations of
single-dish telescopes, particularly as a spatially resolved metric
across the full beam. This provides an important motivation for
extending conventional beam and leakage analyses to a Jones-matrix
conditioning framework.

The Ghana Radio Astronomy Observatory (GRAO) 32-m telescope provides an
important single-dish and very long baseline interferometry facility for
radio astronomy in Africa. Its increasing use for spectral-line,
continuum, pulsar, and Very Long Baseline Interferometry (VLBI) observations creates a corresponding need
for a quantitative understanding of its instrumental polarisation
response. In particular, accurate polarimetric observations require
knowledge of how the antenna response varies with direction across the
primary beam, rather than relying solely on an on-axis calibration or a
scalar beam description.

Previous electromagnetic studies of the GRAO 32-m telescope established
important characteristics of its optical configuration and total-power
performance at 5 and 6.7~GHz, including beamwidth, aperture efficiency,
sidelobe behaviour, and the effects of the shaped reflector,
beam-waveguide, feed illumination, and mechanical tolerances
\citep{venter2018electromagnetic, venter2017electromagnetic}. These
studies provide an important electromagnetic foundation for the present
work. However, a full direction-dependent polarimetric characterisation
of the GRAO 32-m beam has not previously been reported. In particular,
the complex Jones response, the corresponding Mueller response and
Stokes leakage, beam squint, and the conditioning of polarisation
recovery have not been examined together across the primary beam. This
leaves an important gap in assessing the intrinsic polarimetric
performance of the telescope.

In this paper, we present a comprehensive polarisation characterisation
of the GRAO 32-m telescope at 5 and 6.7~GHz. Using high-fidelity
electromagnetic simulations, we derive the direction-dependent Jones
matrix of the antenna and transform it into the corresponding Mueller
representation. We quantify co- and cross-polar beam responses, beam
squint, instrumental Stokes leakage, and the IXR across the primary beam. By combining
field-level, Stokes-level, and conditioning-based metrics, the study
establishes an intrinsic electromagnetic baseline for the polarimetric
response of the GRAO 32-m telescope and provides a framework for its
subsequent observational calibration and validation.

The structure of the paper is as follows. Section~\ref{sec:telescope_em_}
describes the GRAO 32-m telescope and the electromagnetic model adopted
for the simulations. Section~\ref{sec:pol_theory_metrics} establishes
the theoretical framework for the Jones and Mueller representations and
defines the polarimetric performance metrics used in this study.
Section~\ref{sec:data-prep} describes the numerical processing applied
to the simulated electromagnetic fields. The resulting beam,
instrumental leakage, beam squint, and IXR distributions are presented
and analysed in Section~\ref{sec:R&A}. Section~\ref{sec:discussion}
discusses the implications of the results for polarimetric observations
and VLBI applications, while Section~\ref{sec:conclusion} summarises
the main findings and directions for future observational validation.

\section{GRAO 32-m Instrument and Simulation Framework}
\label{sec:telescope_em_}

\subsection{GRAO 32-m Telescope}
\label{subsec:grao_telescope}

The GRAO hosts a fully steerable 32-m single-dish radio telescope located at Kutunse, Ghana,
approximately $25$~km north-west of Accra. The telescope forms part of the Ghana Space Science and Technology Institute (GSSTI) and serves both as an independent single-dish facility and as a station within
the African Very Long Baseline Interferometry (AVN) network. Its location at approximately $5.75^\circ$~N latitude and $0.31^\circ$~W longitude provides access to a large fraction of the celestial sphere, making the telescope suitable
for a broad range of radio astronomical observations and for
participation in global VLBI observations.

The 32-m antenna was originally constructed as a telecommunications ground station and was subsequently converted into a radio telescope through a programme of structural, mechanical, radio-frequency, control, software, and timing-system upgrades. The conversion was undertaken to repurpose the existing infrastructure for astronomical
observations and VLBI operations. Engineering work
included refurbishment of the antenna structure, replacement of
critical azimuth and drive components, reconstruction of the
subreflector support and quad-leg assembly, improvements to antenna centring and pointing, and upgrades to the receiver and timing
systems. Photogrammetric measurements and microwave holography were
also employed during the engineering programme to assess subreflector
alignment and primary-reflector deformation \citep{Nsor2024}. These
activities provide the engineering basis for the telescope geometry
adopted in the electromagnetic model considered in this study.

The telescope employs a shaped dual-reflector Cassegrain optical
configuration combined with a beam-waveguide (BWG) system. The primary
reflector has a diameter of $32$~m and an $f/D$ ratio of approximately
$0.32$. Radiation reflected by the primary is redirected by a
subreflector of approximately $2.90$~m diameter and subsequently
propagates through a sequence of four beam-waveguide mirrors,
comprising two concave and two flat mirrors, before reaching the
receiver cabin located approximately $20$~m below the primary
reflector vertex \citep{ProvenAdzri2026}. The Cassegrain focus is
located approximately $0.84$~m above the primary-reflector vertex.
The shaped reflector geometry is intended to provide the required
combination of aperture efficiency, spillover control, and sidelobe
performance, while the beam-waveguide arrangement allows the receiver
system to remain stationary as the antenna tracks astronomical
sources.

The primary reflector is constructed from segmented panels and has a
specified surface accuracy of approximately $1.88$~mm RMS. The antenna is supported by an alt-azimuth
wheel-and-track mounting system, with an elevation range of
approximately $7^\circ$--$90^\circ$ and an azimuthal range of about
$300^\circ$. The telescope has undergone substantial mechanical
refurbishment as part of its conversion from telecommunications to
radio astronomy, including work on the reflector support structure,
subreflector support, azimuth bearing, drive system, and antenna
control infrastructure.

The GRAO receiver systems operate in the C band, with principal
observing configurations centred near $5.0$ and $6.7$~GHz. The
$5$~GHz system supports continuum and related observations, while the
$6.7$~GHz system is particularly important for observations of the
$6668.518$~MHz Class~II methanol maser transition
\citep{AnsahNarh2026Maser}. The receiver architecture
provides dual circular-polarisation capability, while the telescope
optics and electromagnetic fields are modelled in a coordinate system
defined by the reflector geometry. These two observing frequencies
therefore provide representative cases for investigating the
direction-dependent beam response of the telescope.

\begin{figure}
\begin{minipage}[H]{\linewidth}
\centering
\includegraphics[width=\textwidth]{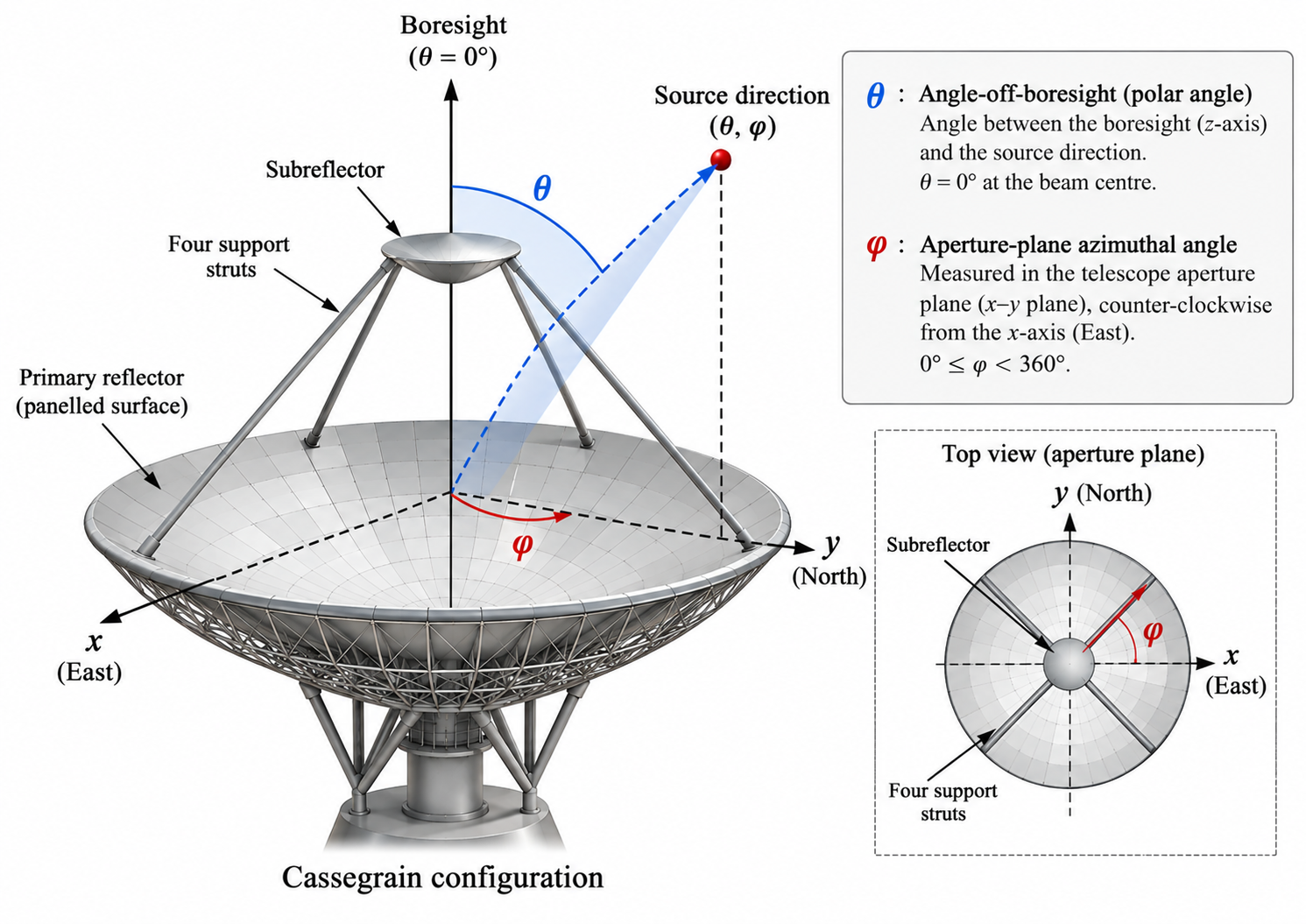}
\end{minipage}
\caption{Geometrical configuration of the GRAO 32-m telescope used in
the electromagnetic simulations. The telescope employs a shaped
Cassegrain dual-reflector system with a beam-waveguide optical path.
The primary reflector, subreflector, and four support struts are
shown. The angle $\theta$ denotes the angular displacement from
boresight, measured between the boresight ($z$-axis) and the source
direction, with $\theta=0^\circ$ corresponding to the beam centre.
The azimuthal angle $\phi$ is measured in the telescope aperture
$(x$--$y$) plane from the $x$-axis and spans $0^\circ$--$360^\circ$.
Each observation direction is therefore specified by the pair
$(\theta,\phi)$, defining the angular coordinate system used for the
electromagnetic beam simulations.}
\label{fig:geometric-sampl}
\end{figure}

\subsection{Electromagnetic Model and GRASP Configuration}
\label{subsec:grasp_model}

The electromagnetic beam of the GRAO 32-m telescope was modelled using
the General Reflector Antenna Software Package (\texttt{GRASP}). The
model represents the principal electromagnetic features of the
telescope optical system, including the shaped primary reflector,
subreflector, beam-waveguide mirrors, feed illumination, and
representative support structures. The objective of the model is to
calculate the direction-dependent far-field response of the adopted
telescope configuration at $5.0$ and $6.7$~GHz. The resulting
electromagnetic fields form the input to the polarimetric formulation
developed subsequently in Section~\ref{sec:pol_theory_metrics}.

\subsubsection{Reflector Geometry and Surface Representation}

The primary and secondary reflectors were represented using the
shaped reflector geometries of the GRAO optical system. The primary
reflector diameter, subreflector dimensions, relative reflector
positions, and beam-waveguide geometry were retained according to the
adopted telescope configuration described in
Section~\ref{subsec:grao_telescope}. The reflecting surfaces were
represented using a segmented computational mesh appropriate for
physical-optics analysis of the electrically large reflector system.

The computational surface resolution was selected to resolve the
electromagnetic variation across the reflector surfaces at the highest
frequency considered. The \texttt{GRASP} surface discretisation was therefore
defined with respect to the electromagnetic wavelength rather than
the angular beam-grid spacing used subsequently for the far-field
analysis. The reflector surface model incorporates the specified
primary-reflector RMS surface accuracy of approximately $1.88$~mm,
which represents the available engineering characterisation of the
GRAO reflector \citep{ProvenAdzri2026, Nsor2024}.

% IMPORTANT: insert the actual GRASP mesh criterion/value here if
% available from the model files or simulation documentation.
% Do not state a numerical mesh size unless it is known.

\subsubsection{Feed Illumination}

The electromagnetic excitation of the optical system was defined by
the feed model used in the GRAO \texttt{GRASP} configuration. The feed
illumination determines the amplitude and phase distribution incident
on the subreflector and consequently influences the illumination of the
primary reflector, spillover, sidelobe structure, and the resulting
far-field response.

The adopted feed excitation was retained consistently for both
operating frequencies so that the resulting beam differences reflect
the frequency-dependent electromagnetic response of the telescope
rather than changes introduced by the subsequent beam-processing
procedure. The feed illumination was propagated through the complete
reflector and beam-waveguide geometry before calculation of the
far-field response.

% IMPORTANT: the exact provenance of the feed pattern must be stated
% here if known:
% (i) measured feed pattern,
% (ii) analytically defined feed pattern,
% (iii) separately simulated feed model, or
% (iv) GRASP feed model supplied as part of the telescope model.
%
% Do not label it "measured" or "analytic" unless this is documented.

\subsubsection{Support Structures and Electromagnetic Blockage}

The model includes the principal support structures associated with
the reflector and subreflector assembly. In particular, the
subreflector support and four primary support struts are retained in
the electromagnetic geometry because these structures occupy part of
the optical path and can introduce blockage and diffraction.

Their inclusion is important for a realistic description of the
far-field beam because structural elements can produce azimuthally
dependent contributions to the sidelobe pattern and can modify the
off-axis electromagnetic response. The support structures were
represented according to their nominal telescope geometry rather than
being replaced by an ideal rotationally symmetric reflector model.
This treatment is consistent with previous electromagnetic
characterisations of the GRAO antenna, in which the non-axisymmetric
beam-waveguide and support configuration contributed to asymmetric
beam structure \citep{AnsahNarh2026FIR, venter2018electromagnetic}.

\subsubsection{Surface Accuracy and Alignment Assumptions}
\label{subsec:surface_alignment}

The electromagnetic model represents the nominal GRAO optical
configuration using the adopted shaped reflector surfaces rather than
an idealised mathematical paraboloid. The primary reflector,
subreflector, beam-waveguide mirrors, feed, and support structures are
retained in their nominal geometrical configuration.

The reference simulations assume nominal alignment of the primary
reflector, subreflector, beam-waveguide mirrors, and feed. No additional
surface deformations or mechanical alignment offsets are introduced in
the simulations. The resulting electromagnetic response therefore
corresponds to the specified nominal configuration used throughout the
analysis.

\subsubsection{Physical Optics and Physical Theory of Diffraction}

The electromagnetic calculations employ the physical optics (PO)
formulation for the induced currents on the conducting reflector
surfaces, together with the physical theory of diffraction (PTD) to
account for diffraction contributions from reflector edges and other
electromagnetically significant discontinuities. This PO--PTD
formulation has previously been applied to electromagnetic modelling
of the GRAO 32-m reflector system \citep{AnsahNarh2026FIR, venter2018electromagnetic}.

Under the physical-optics approximation, the induced surface current
on an illuminated perfectly conducting surface is given by
\begin{equation}
\mathbf{J}_{s}(\mathbf{r}')
=
2\,\hat{\mathbf{n}}
\times
\mathbf{H}_{\mathrm{inc}}(\mathbf{r}'),
\label{eq:surface_current}
\end{equation}
where $\hat{\mathbf{n}}$ is the local outward unit normal to the
reflecting surface and $\mathbf{H}_{\mathrm{inc}}$ is the incident
magnetic field at the surface. The far-field contribution is then
obtained from the radiation integral over the illuminated reflector
surface,
\begin{equation}
\mathbf{E}(\theta,\phi)
\propto
\int_{S}
\mathbf{J}_{s}(\mathbf{r}')
\exp\!\left[
-ik\hat{\mathbf{r}}\cdot\mathbf{r}'
\right]
\,\mathrm{d}S',
\label{eq:far_field}
\end{equation}
where $k=2\pi/\lambda$ is the free-space wavenumber, $\hat{\mathbf r}$
is the observation direction, and $S$ denotes the illuminated
reflector surface.

The PO calculation describes the dominant electromagnetic interaction
with the electrically large reflector surfaces, while PTD augments
the PO solution with diffraction contributions associated with
geometrical discontinuities such as reflector edges. Inclusion of PTD
is particularly relevant when characterising sidelobes and
off-axis response because edge-diffracted fields can contribute to
features that are not represented by an idealised PO-only reflector
model.
The PO and PTD calculations were applied consistently to the adopted
GRAO geometry at both $5.0$ and $6.7$~GHz. The resulting far-field
solutions retain the directional dependence generated by the complete
optical configuration. No scalar beam approximation is introduced at
the electromagnetic modelling stage.

\subsubsection{Angular Sampling of the Electromagnetic Response}

The far-field response was evaluated on a spherical angular grid
centred on the nominal telescope boresight. The polar angle $\theta$
spans $0^\circ$--$3^\circ$ in increments of $0.015^\circ$, while the
azimuthal angle $\phi$ spans the complete $0^\circ$--$360^\circ$
range in $1^\circ$ increments. The angular extent was selected to
include the principal beam and the first sidelobe region at the
frequencies considered.

The fine angular sampling is distinct from the physical surface mesh
used by \texttt{GRASP}. The former controls the sampling of the
computed far-field response, whereas the latter determines the
discretisation of the electromagnetic scattering surfaces. This
distinction is important when assessing the numerical representation
of the beam and avoids conflating reflector-surface discretisation
with subsequent beam-map interpolation.

For each sampled direction, the simulation produces the four complex
field responses associated with the two orthogonal feed excitations
and the two orthogonal far-field components. These quantities are
retained as the electromagnetic simulation output and are subsequently
processed using the polarisation formalism described in
Section~\ref{sec:pol_theory_metrics}.

\section{Polarisation Theory and Metrics}
\label{sec:pol_theory_metrics}

The polarimetric response of a radio telescope is fundamentally a vector transformation that maps the incident electromagnetic field to measured voltages at the receiver terminals. For a single-dish antenna, this transformation is direction-dependent, reflecting the combined effects of feed response, reflector geometry, and propagation through the optical system \citep{van2026introduction}. A rigorous description therefore, requires a formulation that preserves both amplitude and phase information at the field level before propagating to power-based observables.

\subsection{Jones matrix beam representation}
\label{subsec:jones_representation}

At each direction $(\theta,\phi)$ on the sky, the antenna response may be represented by a complex $2\times2$ Jones matrix $\mathbf{J}(\theta,\phi)$, which acts linearly on the incident electric field vector according to
\begin{equation}
\mathbf{E}_{\mathrm{out}}(\theta,\phi)
=
\mathbf{J}(\theta,\phi)\,
\mathbf{E}_{\mathrm{in}}(\theta,\phi),
\label{eq:jones_transform}
\end{equation}
where the Jones matrix is given explicitly by
\begin{equation}
\mathbf{J}(\theta,\phi)
=
\begin{pmatrix}
J_{qh} & J_{qv} \\
J_{ph} & J_{pv}
\end{pmatrix}.
\label{eq:jones_definition}
\end{equation}
Here, the subscripts denote the response of the nominally horizontal $(q)$ and vertical $(p)$ feed channels to horizontally $(h)$ and vertically $(v)$ polarised incident plane waves. The diagonal elements $J_{qh}$ and $J_{pv}$ correspond to the co-polar voltage patterns, while the off-diagonal elements $J_{qv}$ and $J_{ph}$ quantify cross-polar coupling induced by departures from ideal symmetry in the antenna optics and feed system.

Because $\mathbf{J}(\theta,\phi)$ is complex and direction dependent, it encodes not only differential gain but also relative phase delays between polarisation channels. These phase terms are essential for describing leakage between linear and circular polarisation states and cannot be recovered from power-only beam models. In the limit of an ideal, perfectly symmetric system, the Jones matrix reduces to a scalar gain multiplied by the identity matrix, and all cross-polar terms vanish. Real antennas, however, depart from this ideal, particularly away from boresight, necessitating the full matrix treatment described by Eq.~\eqref{eq:jones_transform}.

\subsection{Mueller matrix formulation for a single dish}
\label{subsec:mueller_formulation}

Astronomical radiation is, in general, partially coherent, and polarimetric measurements are therefore most naturally expressed in terms of Stokes parameters rather than instantaneous electric fields. The link between the coherent Jones formalism and the incoherent Stokes description is provided by the coherency matrix
\begin{equation}
\mathbf{C}
=
\langle \mathbf{E}\mathbf{E}^{\dagger} \rangle,
\label{eq:coherency_matrix}
\end{equation}
where angle brackets denote a time average over intervals long compared to the coherence time of the radiation. Under the action of a Jones system, the coherency matrix transforms as
\begin{equation}
\mathbf{C}_{\mathrm{out}}
=
\mathbf{J}\,
\mathbf{C}_{\mathrm{in}}\,
\mathbf{J}^{\dagger}.
\label{eq:coherency_transform}
\end{equation}

The Stokes parameters may be written as linear combinations of the coherency matrix elements using the Pauli matrices $\boldsymbol{\sigma}_i$,
\begin{equation}
S_i
=
\frac{1}{2}
\operatorname{Tr}
\left(
\boldsymbol{\sigma}_i \mathbf{C}
\right),
\quad i = 0,1,2,3,
\label{eq:stokes_definition}
\end{equation}
where $\boldsymbol{\sigma}_0$ is the identity matrix and $\boldsymbol{\sigma}_{1,2,3}$ form a basis for the space of Hermitian $2\times2$ matrices \citep{fano1957description}. Substituting Eq.~\eqref{eq:coherency_transform} into Eq.~\eqref{eq:stokes_definition} yields a linear relation between the input and output Stokes vectors,
\begin{equation}
\mathbf{S}_{\mathrm{out}}(\theta,\phi)
=
\mathbf{M}(\theta,\phi)\,
\mathbf{S}_{\mathrm{in}},
\label{eq:mueller_transform}
\end{equation}
where the direction-dependent Mueller matrix elements are given by
\begin{equation}
M_{ij}(\theta,\phi)
=
\frac{1}{2}
\operatorname{Tr}
\left(
\boldsymbol{\sigma}_i
\mathbf{J}(\theta,\phi)
\boldsymbol{\sigma}_j
\mathbf{J}^{\dagger}(\theta,\phi)
\right),
\quad i,j = 0,\ldots,3.
\label{eq:mueller_elements}
\end{equation}

\noindent
Equation~\eqref{eq:mueller_elements} provides an explicit and exact mapping from the Jones beam to the corresponding Mueller beam and is valid for arbitrary, non-ideal antenna responses. For clarity and to facilitate later physical interpretation, it is useful to write the Mueller matrix explicitly in terms of the complex, direction-dependent Jones matrix elements that describe the antenna voltage response. At each sky direction $(\theta,\phi)$, the corresponding Mueller matrix $\mathbf{M}(\theta,\phi)$ may be written in full as
\begin{equation}
\mathbf{M}(\theta,\phi) =
\begin{pmatrix}
M_{00} & M_{01} & M_{02} & M_{03} \\
M_{10} & M_{11} & M_{12} & M_{13} \\
M_{20} & M_{21} & M_{22} & M_{23} \\
M_{30} & M_{31} & M_{32} & M_{33}
\end{pmatrix},
\label{eq:mueller_matrix}
\end{equation}
where the individual Mueller matrix elements are given by bilinear combinations of the Jones matrix components as
\begin{equation}
\begin{aligned}
M_{00} &= \tfrac{1}{2}\!\left(|J_{qh}|^{2} + |J_{qv}|^{2} + |J_{ph}|^{2} + |J_{pv}|^{2}\right), \\
M_{01} &= \tfrac{1}{2}\!\left(|J_{qh}|^{2} - |J_{qv}|^{2} + |J_{ph}|^{2} - |J_{pv}|^{2}\right), \\
M_{02} &= \Re\!\left(J_{qh}J_{qv}^{*} + J_{ph}J_{pv}^{*}\right), \\
M_{03} &= \Im\!\left(J_{qh}J_{qv}^{*} + J_{ph}J_{pv}^{*}\right), \\[0.5em]
M_{10} &= \tfrac{1}{2}\!\left(|J_{qh}|^{2} + |J_{qv}|^{2} - |J_{ph}|^{2} - |J_{pv}|^{2}\right), \\
M_{11} &= \tfrac{1}{2}\!\left(|J_{qh}|^{2} - |J_{qv}|^{2} - |J_{ph}|^{2} + |J_{pv}|^{2}\right), \\
M_{12} &= \Re\!\left(J_{qh}J_{qv}^{*} - J_{ph}J_{pv}^{*}\right), \\
M_{13} &= \Im\!\left(J_{qh}J_{qv}^{*} - J_{ph}J_{pv}^{*}\right), \\[0.5em]
M_{20} &= \Re\!\left(J_{qh}J_{ph}^{*} + J_{qv}J_{pv}^{*}\right), \\
M_{21} &= \Re\!\left(J_{qh}J_{ph}^{*} - J_{qv}J_{pv}^{*}\right), \\
M_{22} &= \Re\!\left(J_{qh}J_{pv}^{*} + J_{qv}J_{ph}^{*}\right), \\
M_{23} &= \Im\!\left(J_{qh}J_{pv}^{*} + J_{qv}J_{ph}^{*}\right), \\[0.5em]
M_{30} &= \Im\!\left(J_{qh}J_{ph}^{*} + J_{qv}J_{pv}^{*}\right), \\
M_{31} &= \Im\!\left(J_{qh}J_{ph}^{*} - J_{qv}J_{pv}^{*}\right), \\
M_{32} &= \Im\!\left(J_{qh}J_{pv}^{*} + J_{qv}J_{ph}^{*}\right), \\
M_{33} &= \Re\!\left(J_{qh}J_{pv}^{*} - J_{qv}J_{ph}^{*}\right).
\end{aligned}
\label{eq:mueller_elements_expanded}
\end{equation}

\noindent
Equations~\eqref{eq:mueller_matrix} and~\eqref{eq:mueller_elements_expanded} demonstrate explicitly that all Mueller beam elements arise from coherent interference between co-polar and cross-polar voltage responses. Both the amplitudes and relative phases of the Jones matrix elements contribute to the coupling between Stokes parameters, rather than the Mueller response being determined by power terms alone \citep{born2013principles}.

The first row of the Mueller matrix describes how the incident Stokes parameters contribute to the measured total-power beam. In particular, $M_{00}$ corresponds to the Stokes $I$ beam pattern. The elements $M_{10}$ and $M_{20}$ quantify leakage from total intensity into linear polarisation, while $M_{30}$ describes leakage from total intensity into circular polarisation. These terms therefore provide a direct measure of direction-dependent instrumental polarisation induced by the antenna optics.

Beyond the first column, the remaining elements of the Mueller matrix describe how incident linear and circular polarisation states are transformed by the antenna as a function of direction. The diagonal terms \(M_{11}\), \(M_{22}\), and \(M_{33}\) correspond to the effective beam patterns associated with the Stokes \(Q\), \(U\), and \(V\) responses, respectively. In an ideal, perfectly polarimetric system, these terms would be identical to the total-power beam apart from a scalar gain factor. In practice, departures from this ideal behaviour arise from differential illumination, geometric asymmetries, and phase structure in the underlying Jones beams, leading to distinct beam shapes for each Stokes parameter.

The off-diagonal elements encode coupling between different Stokes parameters and therefore represent the full direction-dependent polarimetric response of the antenna. In particular, the terms linking linear polarisation components \((Q \leftrightarrow U)\) and those linking linear and circular polarisation \((Q,U \leftrightarrow V)\) exhibit characteristic quadrupolar and clover-leaf patterns. These spatial symmetries arise from coherent interference between co-polar and cross-polar voltage responses and are a direct manifestation of the phase structure evident in the Jones beams. Such patterns are generic to real reflector systems with non-ideal optical symmetry and have been reported for both single-dish and interferometric antennas \citep{ng2005polarization}.

\subsection{Instrumental polarisation metrics}
\label{subsec:pol_metrics}

Several scalar metrics may be derived from the direction-dependent Mueller matrix to summarise the polarimetric behaviour of the antenna. The half-power beamwidth (HPBW) is defined from the radial profile of $M_{11}(\theta,\phi)$ and characterises the angular resolution of the Stokes $I$ response. Beam squint is quantified as the angular separation between the maxima of the two co-polar power patterns $|J_{qh}|^2$ and $|J_{pv}|^2$, reflecting differential pointing offsets between nominally orthogonal polarisations.

Cross-polarisation is commonly expressed as the ratio of cross-polar to co-polar power,
\begin{equation}
\mathrm{XPR}(\theta,\phi)
=
\frac{|J_{qv}|^2 + |J_{ph}|^2}
{|J_{qh}|^2 + |J_{pv}|^2},
\label{eq:xpr}
\end{equation}
which provides a local measure of polarisation purity at the voltage level. While useful, this metric does not directly quantify the impact on recovered Stokes parameters.

A more observationally relevant quantity is the instrumental polarisation leakage from Stokes $I$ into linear polarisation,
\begin{equation}
\mathrm{IP}(\theta,\phi)
=
\frac{
\sqrt{
M_{21}^2(\theta,\phi) + M_{31}^2(\theta,\phi)
}
}
{M_{11}(\theta,\phi)},
\label{eq:ip_leakage}
\end{equation}
which represents the fractional spurious linear polarisation induced by an unpolarised source. As defined in Eq.~\eqref{eq:ip_leakage}, IP depends on both the magnitudes and relative phases of the Jones elements and can vary significantly across the primary beam.

\subsection{Intrinsic Cross-Polarisation Ratio (IXR)}
\label{subsec:ixr}

While leakage metrics characterise the magnitude of instrumental
polarisation effects, they do not by themselves describe the numerical
stability of recovering the incident polarisation state from the
measured receptor signals. This stability is governed by the
conditioning of the Jones matrix. The condition number of the Jones
matrix is defined as
\begin{equation}
\kappa(\mathbf{J})
=
\frac{\sigma_{\max}}{\sigma_{\min}},
\label{eq:condition_number}
\end{equation}
where $\sigma_{\max}$ and $\sigma_{\min}$ are the maximum and minimum
singular values of $\mathbf{J}$, respectively. Since
$\kappa(\mathbf{J}) \geq 1$, a value of $\kappa=1$ corresponds to a
perfectly conditioned system, whereas increasing values indicate
progressively poorer conditioning and greater sensitivity of the
polarisation reconstruction to measurement noise and errors in the
instrumental response.

The condition number $\kappa$ is therefore already a sufficient
mathematical measure of the conditioning of the Jones matrix. The IXR does not provide an independent
measure of conditioning; rather, it is a one-to-one transformation of
$\kappa$ that expresses the same property in a form more directly
related to polarimetric performance. In the Jones formalism, IXR is
defined as
\begin{equation}
\mathrm{IXR}_{J}
=
\left(
\frac{\kappa(\mathbf{J})+1}
{\kappa(\mathbf{J})-1}
\right)^2.
\label{eq:ixr}
\end{equation}
This formulation relates the condition number to an intrinsic
cross-polarisation ratio and provides a convenient representation of
the orthogonality and relative response of the two polarisation
channels.

The limiting behaviour of IXR provides an intuitive interpretation of
the conditioning. For a perfectly conditioned system,
$\kappa(\mathbf{J}) \rightarrow 1$ and
$\mathrm{IXR}_{J} \rightarrow \infty$, corresponding to perfectly
orthogonal polarisation responses. Conversely, as the Jones matrix
becomes increasingly ill-conditioned, $\kappa(\mathbf{J})$ increases
and $\mathrm{IXR}_{J}$ decreases. In the limiting case where
$\sigma_{\min}\rightarrow0$, $\kappa(\mathbf{J})\rightarrow\infty$
and $\mathrm{IXR}_{J}\rightarrow1$, indicating that the two
polarisation responses have become effectively non-independent and
the incident polarisation cannot be reliably reconstructed.

The principal advantage of reporting IXR rather than $\kappa$ alone is
therefore not that it contains additional mathematical information,
but that it provides a polarimetry-oriented scale that is more readily
interpreted as a cross-polarisation or polarisation-purity metric.
This is particularly useful for highly conditioned systems, where
$\kappa$ may remain numerically close to unity even when the
corresponding change in intrinsic polarisation purity is substantial.
For consistency with common radio-polarimetric practice, IXR is also
conveniently expressed in decibels as
\begin{equation}
\mathrm{IXR}_{J,\mathrm{dB}}
=
10\log_{10}\left(\mathrm{IXR}_{J}\right).
\label{eq:ixr_db}
\end{equation}
Thus, higher IXR values correspond to better intrinsic polarimetric
conditioning and greater tolerance to perturbations in the measured
receptor signals.

An important property of IXR is that it characterises the intrinsic
response of the polarimeter rather than the performance of a particular
calibration algorithm. Because it is derived directly from the
singular values of the Jones matrix, it provides a coordinate-invariant
measure of the separation and relative response of the two
polarisation channels. It therefore describes a fundamental property
of the instrumental response that remains relevant even when a
calibration procedure is applied. A well-conditioned system can be
calibrated more robustly because inversion of its Jones matrix is less
sensitive to measurement noise and uncertainties in the instrumental
model.

For a single-dish telescope, the Jones matrix and hence both
$\kappa$ and IXR are functions of observing direction and frequency,
\begin{equation}
\mathbf{J}=\mathbf{J}(\theta,\phi,\nu),
\end{equation}
where $\theta$ is the angle-off-boresight and $\phi$ is the
aperture-plane azimuth. Consequently, IXR provides a direction- and
frequency-dependent description of the intrinsic polarimetric
conditioning of the telescope. This is particularly relevant to the
present study because off-axis cross-polar coupling, differential
amplitude response, and phase differences can progressively degrade
the conditioning of the polarisation response across the primary beam.
Reporting IXR together with the instrumental-polarisation and beam
metrics therefore provides a more complete characterisation of the
GRAO polarimetric response than any single leakage metric alone
\citep{foster2015intrinsic, carozzi2011fundamental, Tinbergen2005}.

\section{Data Processing}
\label{sec:data-prep}

The simulated direction-dependent Jones voltage patterns from
\texttt{GRASP} were processed numerically to obtain the polarimetric
beam characteristics presented in this work. The processing workflow
comprises input validation, transformation from the linear to circular
polarisation basis, spherical-to-tangent-plane mapping, complex Jones
field interpolation, and subsequent construction of the direction-
dependent Mueller matrices and polarimetric performance metrics.

The \texttt{GRASP} outputs were imported directly from the native
\texttt{MATLAB} binary format into a \texttt{Python}-based analysis environment using
standard scientific libraries. The Jones voltage patterns were retained
as complex-valued arrays sampled on the native spherical
$(\theta,\phi)$ grid defined by the electromagnetic simulation. Before
the subsequent polarimetric processing, the input arrays and angular
coordinates were checked for consistency, including the expected
angular ranges, array dimensions, and complex-valued Jones components.
The co-polar voltage responses were verified to peak at boresight, as
expected for the aligned reflector configuration. The amplitude and
phase distributions were also inspected over the native angular grid
for continuity and spatial consistency. These checks provide a direct
assessment of the native simulated beam before any coordinate
transformation or interpolation is performed. Apparent phase jumps
associated with the $2\pi$ representation of phase were distinguished
from discontinuities in the underlying complex Jones voltage. Where
appropriate, the expected approximate symmetries of the optical
configuration were also inspected qualitatively, while recognising
that departures from perfect symmetry can arise intrinsically from the
shaped-reflector and beam-waveguide geometry.

Although the \texttt{GRASP} simulations provide the antenna voltage
response in a linear polarisation basis, the GRAO C-band receiver
system operates in a circular polarisation basis. Circular
polarisation is also widely used in radio astronomy and very
long-baseline interferometry, and is particularly relevant to
observations of sources such as pulsars and masers. The simulated
Jones matrices were therefore transformed from the linear $(H,V)$
basis defined in Eq.~\eqref{eq:jones_definition} into the circular
$(R,L)$ basis before construction of the Mueller matrices and
evaluation of the polarimetric metrics.

The circular-basis Jones matrix was obtained via the unitary basis
transformation
\begin{equation}
J_{\mathrm{circ}} = U J_{\mathrm{lin}} U^{\dagger},
\qquad
U = \frac{1}{\sqrt{2}}
\begin{pmatrix}
1 & i \\
1 & -i
\end{pmatrix},
\end{equation}
which yields the explicit relations
\begin{align}
J_{RR} &= \tfrac{1}{2}(J_{qh}+J_{pv})
       + \tfrac{i}{2}(J_{qv}-J_{ph}), \\
J_{RL} &= \tfrac{1}{2}(J_{qh}-J_{pv})
       + \tfrac{i}{2}(J_{qv}+J_{ph}), \\
J_{LR} &= \tfrac{1}{2}(J_{qh}-J_{pv})
       - \tfrac{i}{2}(J_{qv}+J_{ph}), \\
J_{LL} &= \tfrac{1}{2}(J_{qh}+J_{pv})
       - \tfrac{i}{2}(J_{qv}-J_{ph}).
\end{align}
This transformation was applied pointwise to the complete angular
grid, producing the direction-dependent Jones operator in the circular
basis used throughout the subsequent polarimetric analysis.

\subsection{Tangent-Plane Mapping and Interpolation}
\label{subsec:interpolation}

The Jones voltage patterns produced by \texttt{GRASP} are sampled on a
spherical coordinate grid parameterised by the polar angle $\theta$,
which measures angular offset from boresight, and the azimuthal angle
$\phi$, defined in the aperture $(x$--$y)$ plane. For subsequent
visualisation and pixel-based beam analysis, the spherical sampling was
mapped onto a two-dimensional tangent plane centred on the telescope
boresight. Over the primary-beam field considered in this work, the
small-angle approximation is adequate, and the angular coordinates
$(x,y)$ were defined as
\begin{equation}
x = \theta \cos\phi, \qquad
y = \theta \sin\phi ,
\end{equation}
with inverse relations
\begin{equation}
\theta = \sqrt{x^{2}+y^{2}}, \qquad
\phi = \operatorname{atan2}(y,x),
\end{equation}
where $\phi$ is mapped onto $[0^\circ,360^\circ)$.

A regular Cartesian grid spanning the required field of view was then
constructed in the $(x,y)$ plane. The target grid used in the analysis
covers a $1^\circ\times1^\circ$ field of view with $512\times512$
pixels, corresponding to a spatial sampling of approximately
\begin{equation}
\Delta x = \Delta y
\simeq \frac{1^\circ}{512}
=0.001953^\circ
\simeq 7.03~\mathrm{arcsec}.
\end{equation}
This target grid is a numerical resampling of the native
electromagnetic solution; it does not represent an increase in the
intrinsic angular information contained in the \texttt{GRASP}
simulation.

The interpolation was performed directly on the complex Jones
voltages in the circular basis using linear interpolation. In
particular, the complex Jones components were retained in their
Cartesian complex representation during interpolation rather than
being decomposed into amplitude and wrapped phase. Thus, the
interpolation does not require phase unwrapping and does not
interpolate a phase quantity containing artificial $2\pi$ discontinuities.
For a complex Jones component $J$, the interpolation therefore acts on
the complex field itself,
\begin{equation}
J(x,y) \simeq \sum_{n} w_n(x,y)J_n,
\end{equation}
where $J_n$ are the neighbouring native \texttt{GRASP} complex samples and
$w_n$ are the interpolation weights. The subsequent Mueller and
polarimetric quantities are calculated from these interpolated Jones
operators rather than interpolating the derived polarimetric metrics.

The native \texttt{GRASP} angular sampling in $\theta$ is $0.015^\circ$
($54$~arcsec). At 5~GHz, the characteristic angular scale
corresponding to $\lambda/(2D)$ for the 32-m aperture is
\begin{equation}
\frac{\lambda}{2D}
=
\frac{0.06}{2\times32}
\simeq 0.0009375~\mathrm{rad}
\simeq 0.0537^\circ
\simeq 193~\mathrm{arcsec}.
\end{equation}
The native $\theta$ sampling is therefore approximately $3.6$ times
finer than this characteristic scale. Relative to the native
$54$-arcsec sampling, the $7.03$-arcsec target grid is approximately
$7.7$ times finer. These ratios demonstrate that the native
electromagnetic beam is already spatially resolved at the angular
scale considered, while the tangent-plane grid provides a substantially
finer representation for subsequent numerical operations.

The apparent phase discontinuities that occur in phase visualisations
of cross-polar Jones components require particular care in
interpretation. Phase is displayed modulo $2\pi$, so a transition
between phases near $+\pi$ and $-\pi$ can appear as a sharp
discontinuity even when the underlying complex Jones field varies
continuously. Because the present interpolation is applied directly to
the complex Jones voltages, rather than to the displayed wrapped
phase, such modulo-$2\pi$ phase transitions do not constitute evidence
of interpolation-induced phase artefacts. Inspection of the native
beam distributions showed that the spatial structures retained after
resampling are continuous in the underlying complex representation and
are not associated with isolated interpolation-generated features.

As a numerical sensitivity check, the adopted linear interpolation was
also compared with a higher-order cubic interpolation for representative
radial beam cuts. The maximum amplitude difference between the two
interpolation schemes was below $0.1\%$, while the RMS phase difference
was below $0.01$~rad. Propagating the two interpolated Jones fields
through the complete polarimetric analysis resulted in differences of
less than $0.01$~dB in IXR and less than $0.0001\%$ in the derived
leakage. These differences are negligible relative to the
direction-dependent variations of the beam reported in this study and
demonstrate that the reported polarimetric behaviour is not materially
dependent on the selected interpolation order.

No additional anti-aliasing filter or smoothing operation was applied
during the spherical-to-tangent-plane transformation. The operation
constitutes interpolation onto a finer output grid rather than
downsampling of the native electromagnetic solution. Consequently, no
unreported spatial filtering is introduced into the Jones beam. The
native angular sampling and the interpolation-sensitivity comparison
provide the numerical basis for assessing the adequacy of the
resampling procedure.

For clarity and reproducibility, the complete numerical sequence of
input validation, basis transformation, coordinate projection,
complex-field interpolation, and polarimetric metric evaluation is
summarised in Algorithm~\ref{alg:grasp_pipeline}.

\begin{algorithm}
\caption{Processing pipeline for \texttt{GRASP} Jones beams}
\label{alg:grasp_pipeline}
\begin{algorithmic}[1]
\Require Native spherical grid $\{\theta_i,\phi_j\}$ and linear-basis
Jones voltages $J_{qh},J_{qv},J_{ph},J_{pv}$
\Ensure Tangent-plane Jones beams in circular polarisation basis

\State Load \texttt{GRASP} outputs and extract $\theta$, $\phi$, and
complex Jones voltages
\State Verify angular ranges, array dimensions, and complex-valued Jones
components
\State Verify that the co-polar responses peak at $\theta=0$
(boresight)
\State Inspect native amplitude and phase distributions for spatial
continuity and numerical artefacts
\State Construct a Cartesian tangent-plane grid $(x_k,y_l)$ spanning
the specified field of view
\ForAll{$(x_k,y_l)$}
    \State Compute $\theta \gets \sqrt{x_k^2+y_l^2}$
    \State Compute $\phi \gets \operatorname{atan2}(y_l,x_k)$
\EndFor

\State Transform the native linear-basis Jones matrix:
\[
J_{\mathrm{circ}} \gets UJ_{\mathrm{lin}}U^\dagger
\]

\State Interpolate each complex circular-basis Jones component onto the
tangent-plane grid using linear interpolation

\State For a representative subset, repeat interpolation using a
higher-order scheme and compare Jones-field and derived polarimetric
quantities

\State Construct direction-dependent Mueller matrices from the
interpolated Jones operators

\State Evaluate leakage, IXR, and other polarimetric metrics from the
resulting Mueller/Jones quantities
\end{algorithmic}
\end{algorithm}

\subsection{Mueller Matrix Construction and Polarimetric Metrics}
\label{subsec:mueller_metrics}

From the resulting tangent-plane, circular-basis Jones operators, full
direction-dependent Mueller matrices were constructed numerically using
the trace formulation derived in Section~\ref{sec:pol_theory_metrics}.
The computation was vectorised over the full angular grid. The Jones
operators were retained as complex-valued quantities throughout the
calculation, with real and imaginary components handled explicitly in
the numerical implementation.

For each angular direction $(\theta,\phi)$, the singular values required
for the numerical evaluation of the IXR were obtained through the
singular value decomposition (SVD) of the complex $2\times2$ Jones
matrix $\mathbf{J}(\theta,\phi)$,
\begin{equation}
\mathbf{J}
=
\mathbf{U}\boldsymbol{\Sigma}\mathbf{V}^{\dagger},
\label{eq:svd}
\end{equation}
where $\mathbf{U}$ and $\mathbf{V}$ are unitary matrices and
\begin{equation}
\boldsymbol{\Sigma}
=
\operatorname{diag}
\left(\sigma_{\mathrm{max}},\sigma_{\mathrm{min}}\right),
\label{eq:singular_values}
\end{equation}
contains the singular values ordered in descending magnitude.

To ensure numerical robustness in directions where the Jones matrix
becomes nearly singular, a lower bound was imposed on the minimum
singular value,
\begin{equation}
\sigma_{\mathrm{min}}
\leftarrow
\max\left(
\sigma_{\mathrm{min}},
\varepsilon\sigma_{\mathrm{max}}
\right),
\qquad
\varepsilon=10^{-12},
\label{eq:sigma_floor}
\end{equation}
thereby preventing division by zero and loss of floating-point
precision. Condition numbers exceeding $10^{12}$ were clipped to this
threshold prior to evaluation of Eq.~\eqref{eq:ixr}, avoiding numerical
overflow in the IXR calculation for extremely ill-conditioned
directions \citep{bai2021matrix,golub2013matrix,van1996matrix}.

All beam quantities were normalised relative to the peak Stokes-$I$
beam at boresight, enabling leakage terms to be interpreted as
fractional contamination relative to total intensity. Where required
for visualisation or radial averaging, the spherical coordinates
$(\theta,\phi)$ were transformed into direction cosines $(l,m)$ under
the small-angle approximation, facilitating comparison with standard
image-plane representations \citep{thompson2017interferometry}.

From the direction-dependent Mueller matrices, Stokes beam patterns,
instrumental leakage maps, and intrinsic cross-polarisation ratio
distributions were derived deterministically without additional
smoothing or empirical fitting. The resulting products therefore
retain the spatial structure of the simulated direction-dependent
Jones response while avoiding the introduction of empirical corrections
or post-processing filters. These products form the basis of the
quantitative analysis presented in the following section.

\subsection{Jones and Mueller Beam Representations}

\begin{figure*}
\begin{minipage}[H]{\linewidth}
\centering
\includegraphics[width=\textwidth]{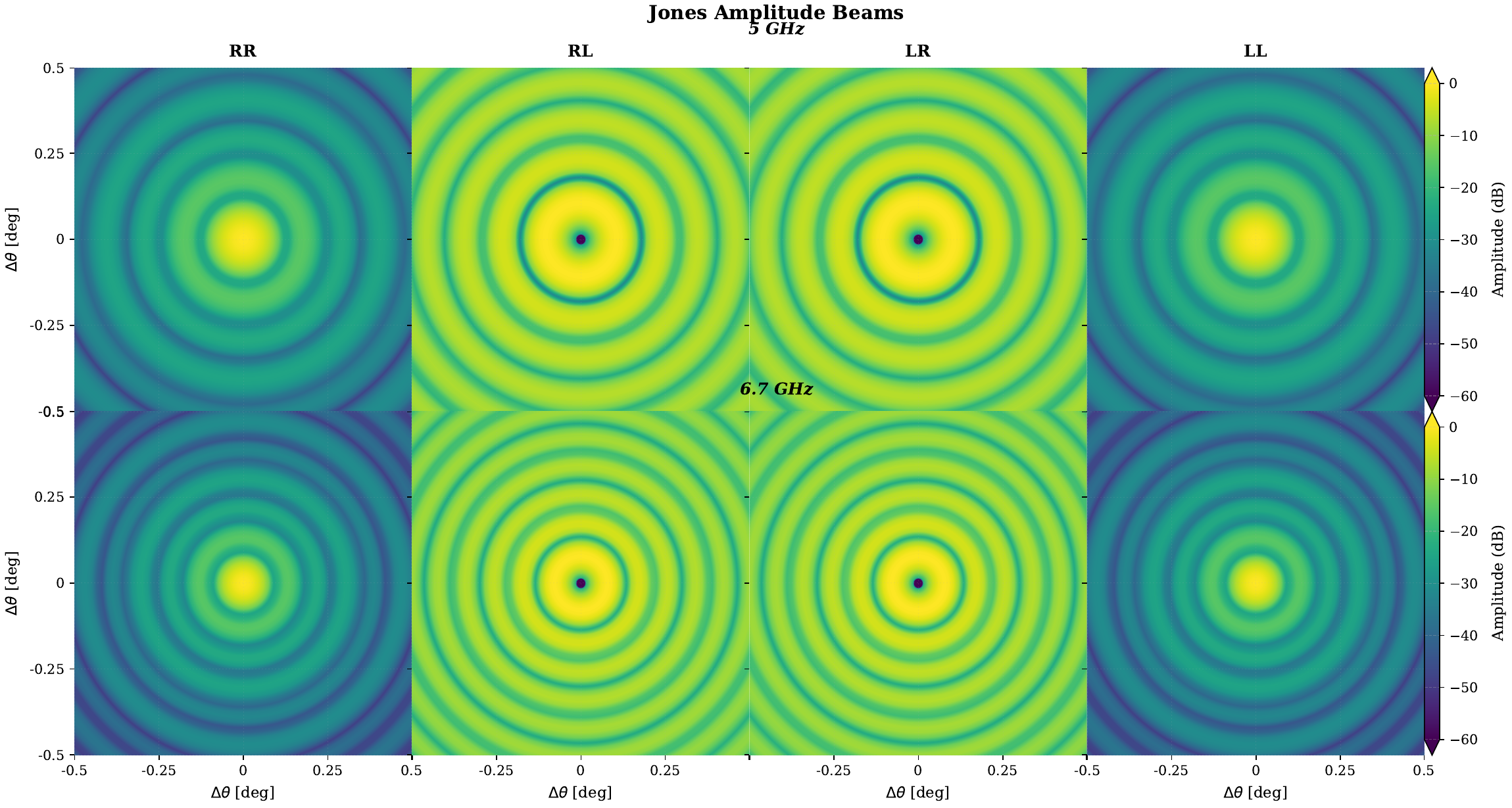} 
\end{minipage}
\caption{Jones amplitude beams of the GRAO 32-m telescope at $5$ GHz (top row) and $6.7$ GHz (bottom row).
The absolute amplitudes of the four circular-polarisation Jones components (RR, RL, LR, LL), normalised to the peak co-polar response and expressed in decibels. The co-polar terms exhibit centrally peaked main beams with concentric sidelobe structure, while the cross-polar terms show central nulls and enhanced off-axis structure. The comparison highlights systematic frequency-dependent changes in sidelobe spacing and amplitude while preserving the overall beam morphology.}
\label{fig:jones_amplitude_beams}
\end{figure*}
%% phase

\begin{figure*}
\begin{minipage}[H]{\linewidth}
\centering
\includegraphics[width=\textwidth]{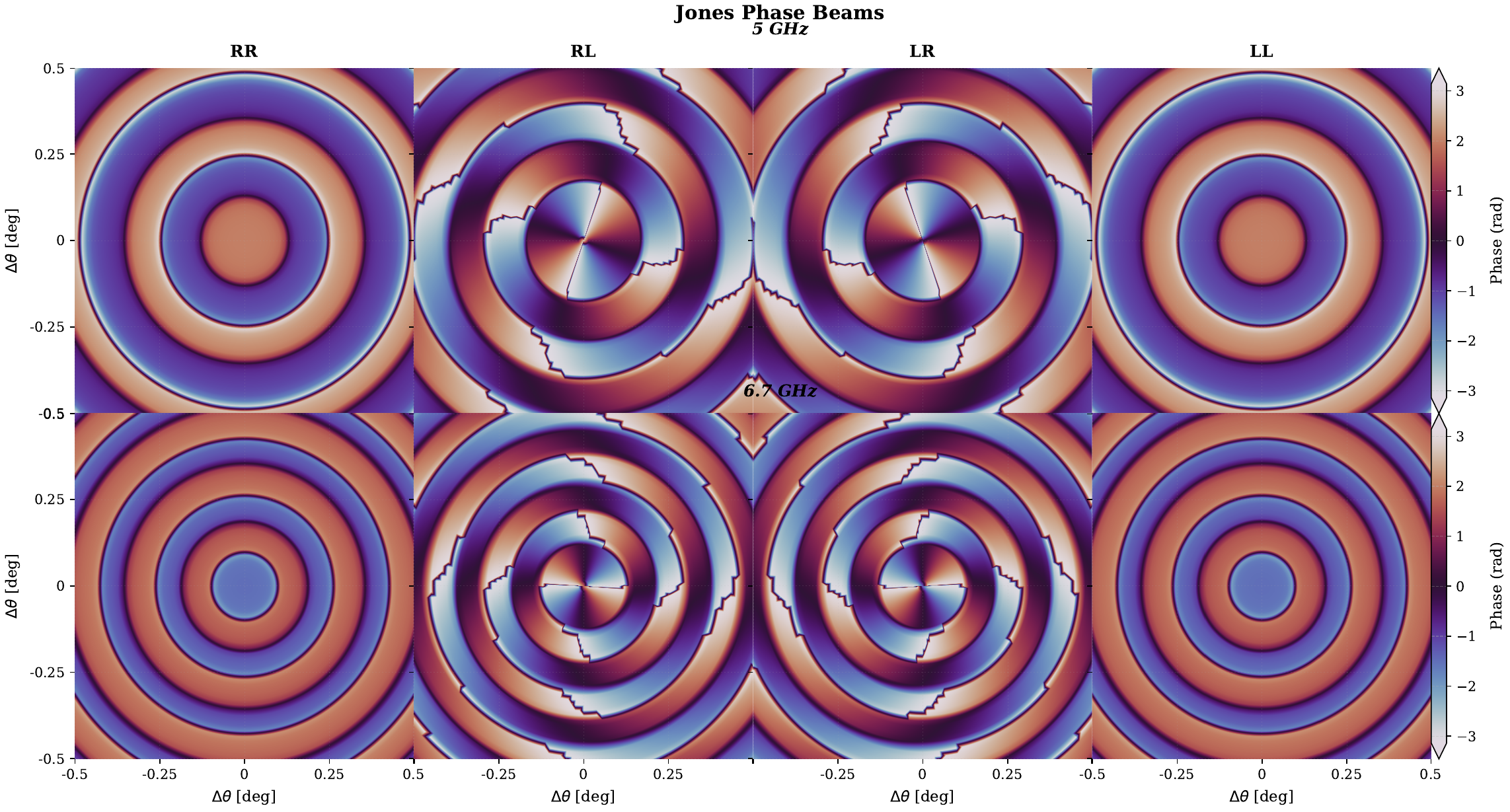} 
\end{minipage}
\caption{Jones phase beams of the GRAO 32-m telescope at $5$ GHz (top row) and $6.7$ GHz (bottom row).
The phase of each circular-polarisation Jones component is shown in radians over the same field of view as Fig.~\ref{fig:jones_amplitude_beams}. The co-polar terms display predominantly radially symmetric phase behaviour near boresight, whereas the cross-polar terms exhibit strong azimuthal phase winding and sharp phase discontinuities. These features are intrinsic to the antenna response and play a central role in direction-dependent polarimetric coupling.}
\label{fig:jones_phase_beams}
\end{figure*}
%% radial cuts

\begin{figure*}
\begin{minipage}[H]{\linewidth}
\centering
\includegraphics[width=\textwidth]{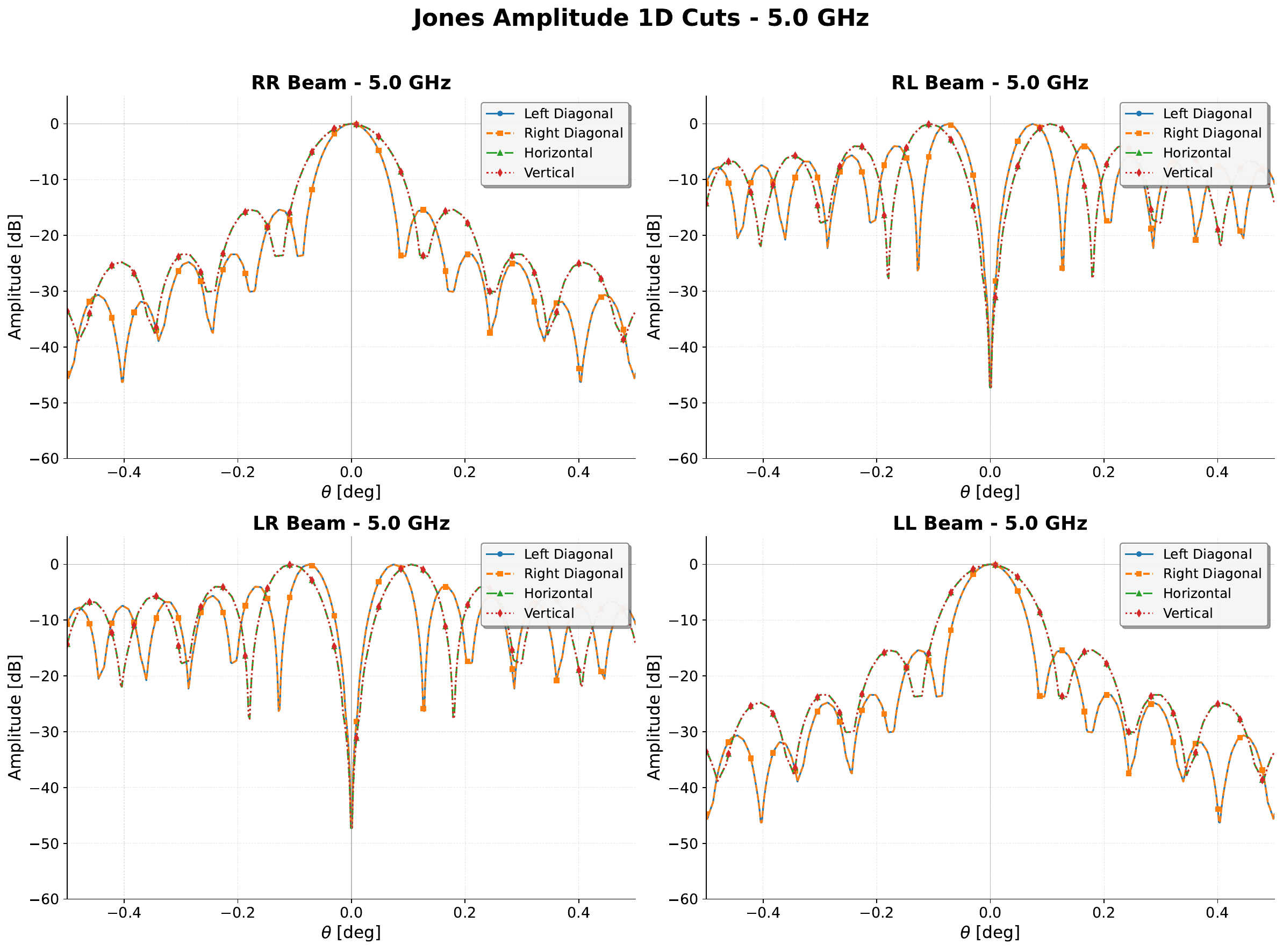} 
\end{minipage}
\caption{One-dimensional radial cuts of the Jones amplitude at $5$ GHz.
Amplitude profiles are shown for the RR, RL, LR, and LL Jones components along the horizontal, vertical, and diagonal directions through the beam centre. The co-polar terms decrease smoothly with angular offset, while the cross-polar terms remain suppressed near boresight and increase more rapidly off-axis. The directional dependence of the cross-polar cuts reflects asymmetries introduced by the antenna optics and feed system.}
\label{fig:jones_amplitude_1Dcuts_5.0GHz}
\end{figure*}

\begin{figure*}
\begin{minipage}[H]{\linewidth}
\centering
\includegraphics[width=\textwidth]{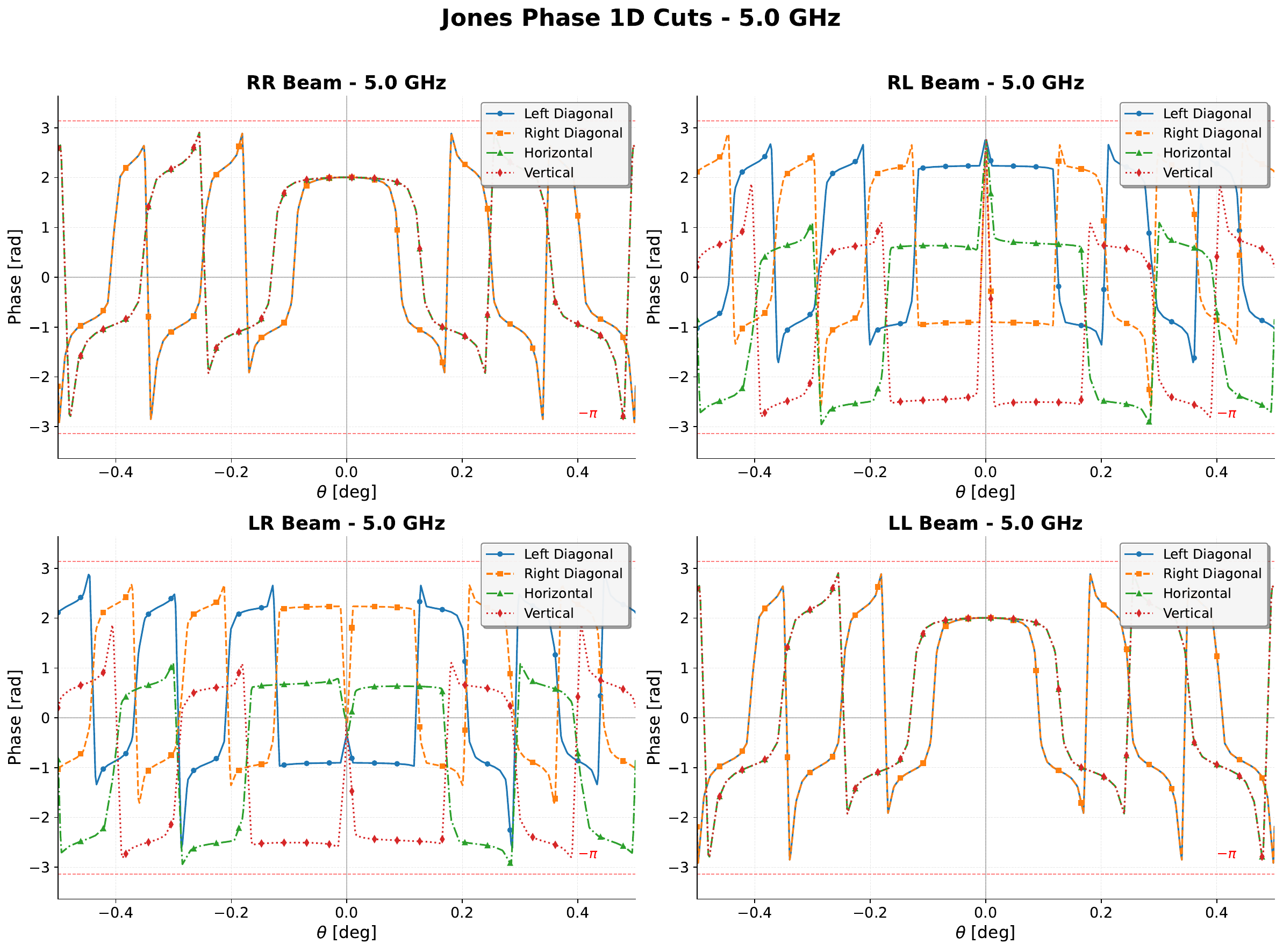} 
\end{minipage}
\caption{One-dimensional radial cuts of the Jones phase at $5$ GHz.
Phase profiles corresponding to the amplitude cuts in Fig.~\ref{fig:jones_amplitude_1Dcuts_5.0GHz} are shown in radians. The cross-polar terms exhibit abrupt phase transitions and wrapping behaviour as a function of angular offset, in contrast to the smoother phase variation of the co-polar responses. These phase features are not captured by power-only beam descriptions and directly contribute to direction-dependent mixing between Stokes parameters.}
\label{fig:jones_phase_1Dcuts_5.0GHz}
\end{figure*}

The field-level structure implied by the direction-dependent Jones matrix is illustrated in Figs.~\ref{fig:jones_amplitude_beams}--\ref{fig:jones_phase_1Dcuts_5.0GHz}, which present the complex voltage response of the antenna in both two-dimensional and one-dimensional forms. These figures provide a compact but comprehensive view of the spatial behaviour of the co- and cross-polar voltage patterns that underpin all subsequent polarimetric effects.

Figure~\ref{fig:jones_amplitude_beams} shows the amplitude of the Jones beams for the four circular-polarisation products at $5$ GHz and $6.7$ GHz. The co-polar terms exhibit centrally peaked responses with concentric sidelobe structure, while the cross-polar terms display markedly different morphologies, including central nulls and enhanced off-axis structure. Although the overall beam shapes are similar between the two frequencies, the relative prominence and spacing of sidelobes change systematically, reflecting the expected frequency scaling of the electromagnetic response. These amplitude patterns demonstrate that departures from an ideal scalar gain occur well within the main beam and therefore cannot be neglected in high-fidelity polarimetric applications.

The corresponding Jones phase distributions are shown in Fig.~\ref{fig:jones_phase_beams}. The co-polar terms exhibit predominantly radially symmetric phase behaviour near boresight, while the cross-polar terms display pronounced azimuthal phase winding and sharp phase discontinuities associated with sign changes in the complex voltage response. These phase structures are intrinsic to the antenna optics and feed geometry and are not artefacts of numerical interpolation. Their presence is particularly significant, as phase gradients and discontinuities directly contribute to interference terms in the Mueller matrix and hence to direction-dependent mixing between Stokes parameters.

To further elucidate the off-axis behaviour of the Jones beams, Fig.~\ref{fig:jones_amplitude_1Dcuts_5.0GHz} presents one-dimensional radial cuts of the Jones amplitude at $5$ GHz, extracted along the principal axes and diagonals of the beam. The co-polar responses decrease smoothly from boresight, while the cross-polar responses remain suppressed near the beam centre and rise more rapidly with angular offset. The close agreement between cuts taken along different directions for the co-polar terms indicates approximate rotational symmetry near boresight, whereas the increased directional variability in the cross-polar cuts highlights the inherently asymmetric nature of the cross-polar response.

Figure~\ref{fig:jones_phase_1Dcuts_5.0GHz} shows the corresponding one-dimensional radial cuts of the Jones phase at $5$ GHz. These plots reveal abrupt phase transitions and wrapping behaviour, particularly in the cross-polar terms, that are not evident from amplitude information alone. Such behaviour underscores the necessity of retaining the full complex Jones representation when assessing polarimetric performance, as power-only descriptions are insufficient to capture the phase-dependent coupling mechanisms that ultimately give rise to instrumental polarisation leakage.
These complex voltage patterns are therefore propagated into Mueller space and quantified in terms of observable polarimetric metrics in the following sections.

\begin{figure*}
\begin{minipage}[H]{\linewidth}
\centering
\includegraphics[width=\textwidth]{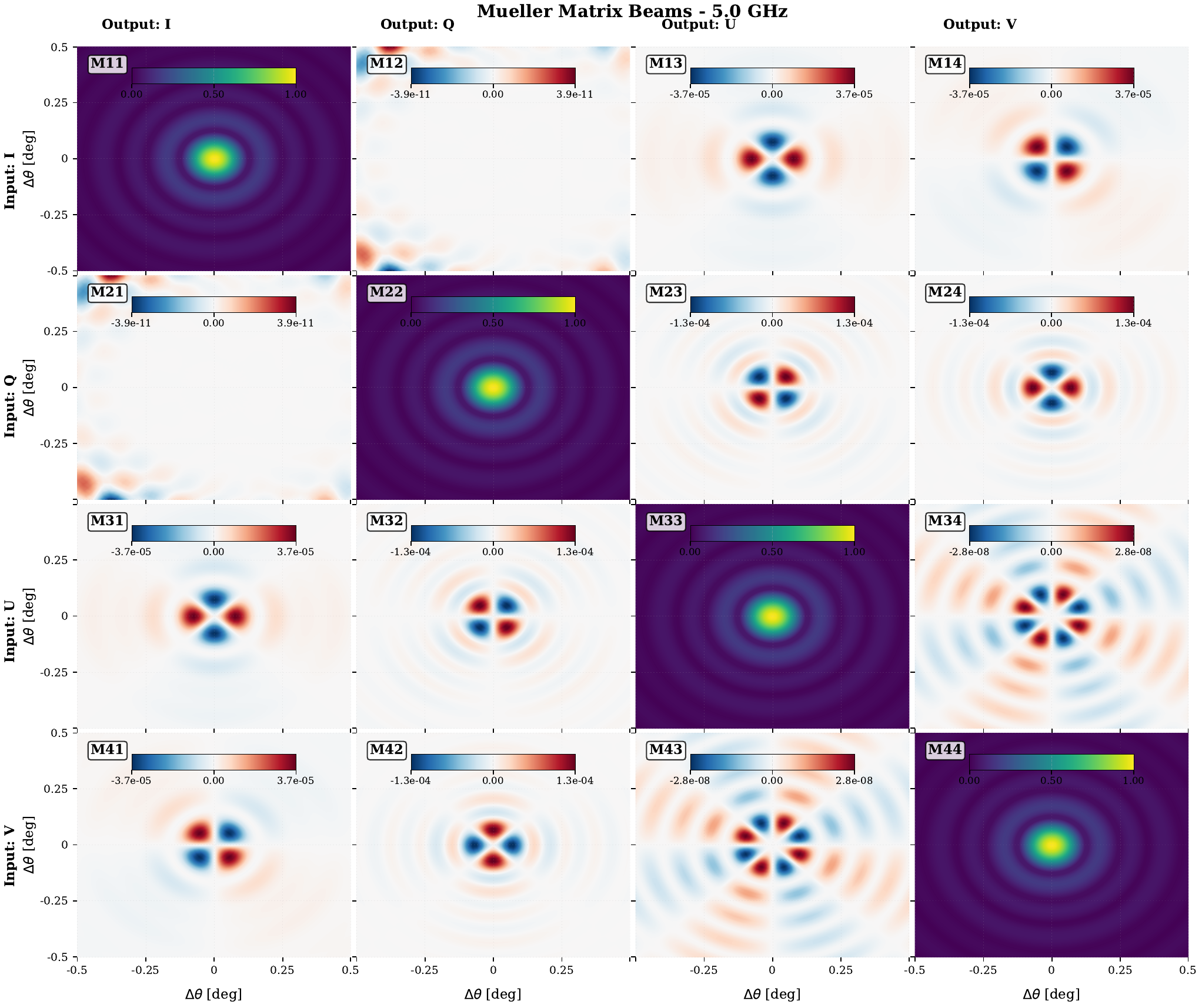} 
\end{minipage}
\caption{Direction-dependent Mueller matrix beams of the GRAO 32-m telescope at $5.0$ GHz.
Each panel shows one element \(M_{ij}(\theta,\phi)\) of the Mueller matrix, arranged such that rows correspond to the input Stokes parameters \((I,Q,U,V)\) and columns correspond to the output Stokes parameters. The diagonal elements \((M_{11}\), \(M_{22}\), \(M_{33}\), \(M_{44})\) represent the effective beam patterns for the corresponding Stokes parameters, while the off-diagonal elements quantify direction-dependent coupling between different polarisation states. All beams are shown over the same angular field of view and are normalised consistently to facilitate comparison of relative structure.}
\label{fig:mueller_matrix_5_0ghz}
\end{figure*}

\begin{figure*}
\begin{minipage}[H]{\linewidth}
\centering
\includegraphics[width=\textwidth]{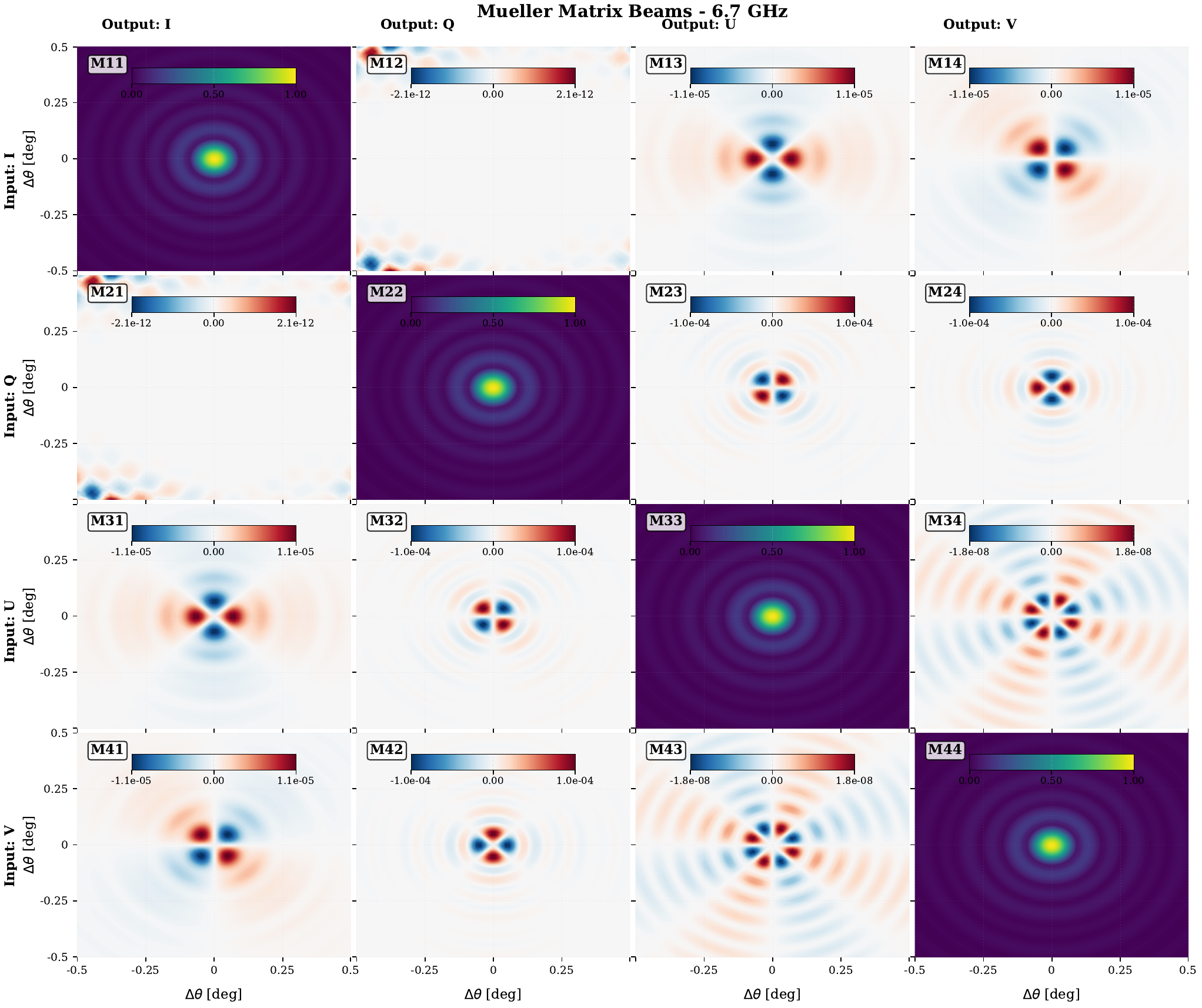} 
\end{minipage}
\caption{Direction-dependent Mueller matrix beams of the GRAO 32-m telescope at $6.7$ GHz.
 Comparison with the $5.0$ GHz case reveals similar qualitative structure across the Mueller elements, with frequency-dependent changes in the relative amplitude and spatial extent of the off-diagonal terms. These variations reflect the frequency dependence of the underlying Jones beams and motivate the quantitative polarimetric analysis presented in later sections.}
\label{fig:mueller_matrix_6_7ghz}
\end{figure*}

Figures~\ref{fig:mueller_matrix_5_0ghz} and~\ref{fig:mueller_matrix_6_7ghz} present the full direction-dependent Mueller matrix beams of the GRAO 32-m telescope at 5.0 and 6.7 GHz, respectively. These figures provide a compact visual representation of the complete polarimetric operator described by Eq.~\eqref{eq:mueller_elements}. While no quantitative assessment is made at this stage, the spatial structure of the Mueller elements highlights the strongly non-scalar nature of the antenna response away from boresight. In particular, the presence of structured off-diagonal terms within and beyond the main beam underscores the necessity of treating instrumental polarisation as a direction-dependent operator rather than as a single, global correction factor.

The Mueller beam representations shown here form the basis for the quantitative analysis presented in subsequent sections. In Section \ref{sec:R&A}, these direction-dependent operators are used to derive Stokes beam properties, instrumental polarisation leakage metrics, and intrinsic cross-polarisation ratios, enabling a systematic assessment of polarimetric performance across the primary beam.

\section{Results and Analysis}\label{sec:R&A}

\begin{figure*}
\begin{subfigure}[t]{\linewidth}
\includegraphics[width=\textwidth]{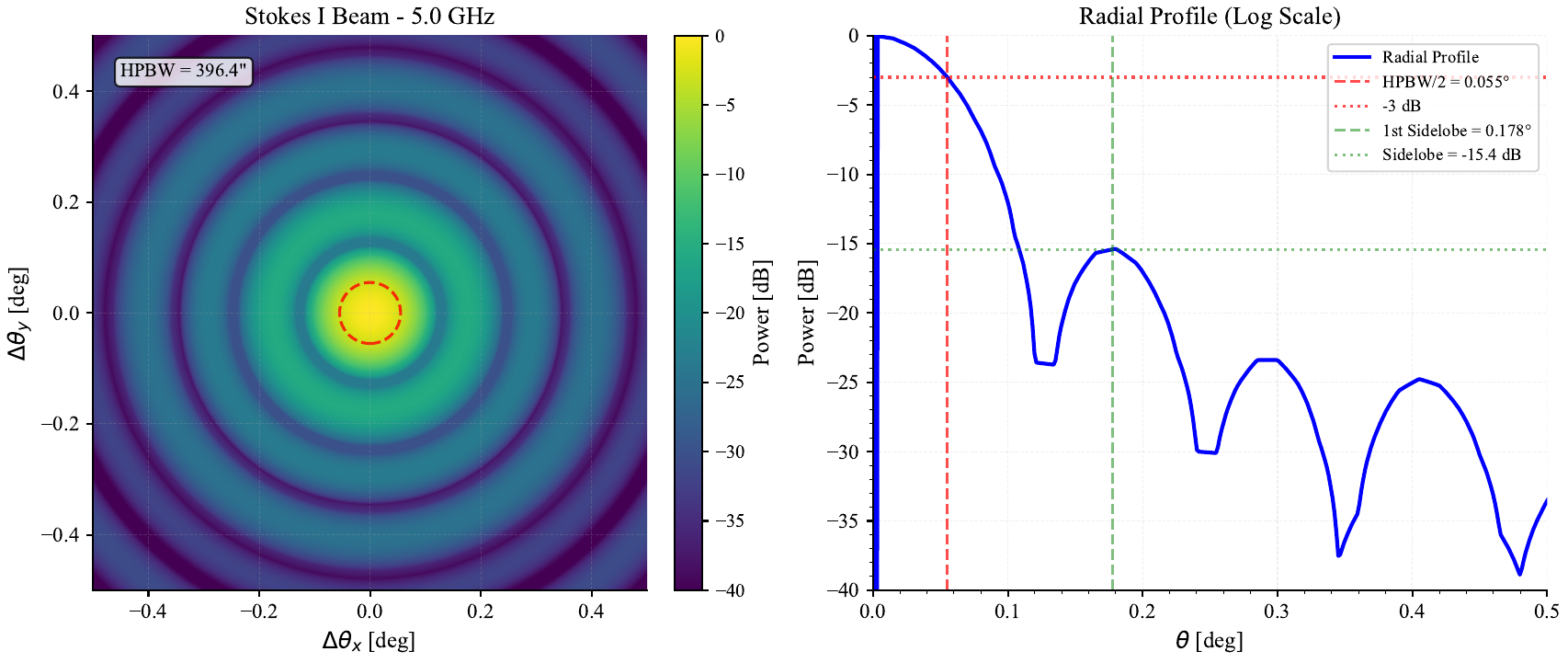}
\caption{}
\label{fig:stokes_I_analysis_5}
\end{subfigure}
\hfill
\begin{subfigure}[t]{\linewidth}
\includegraphics[width=\textwidth]{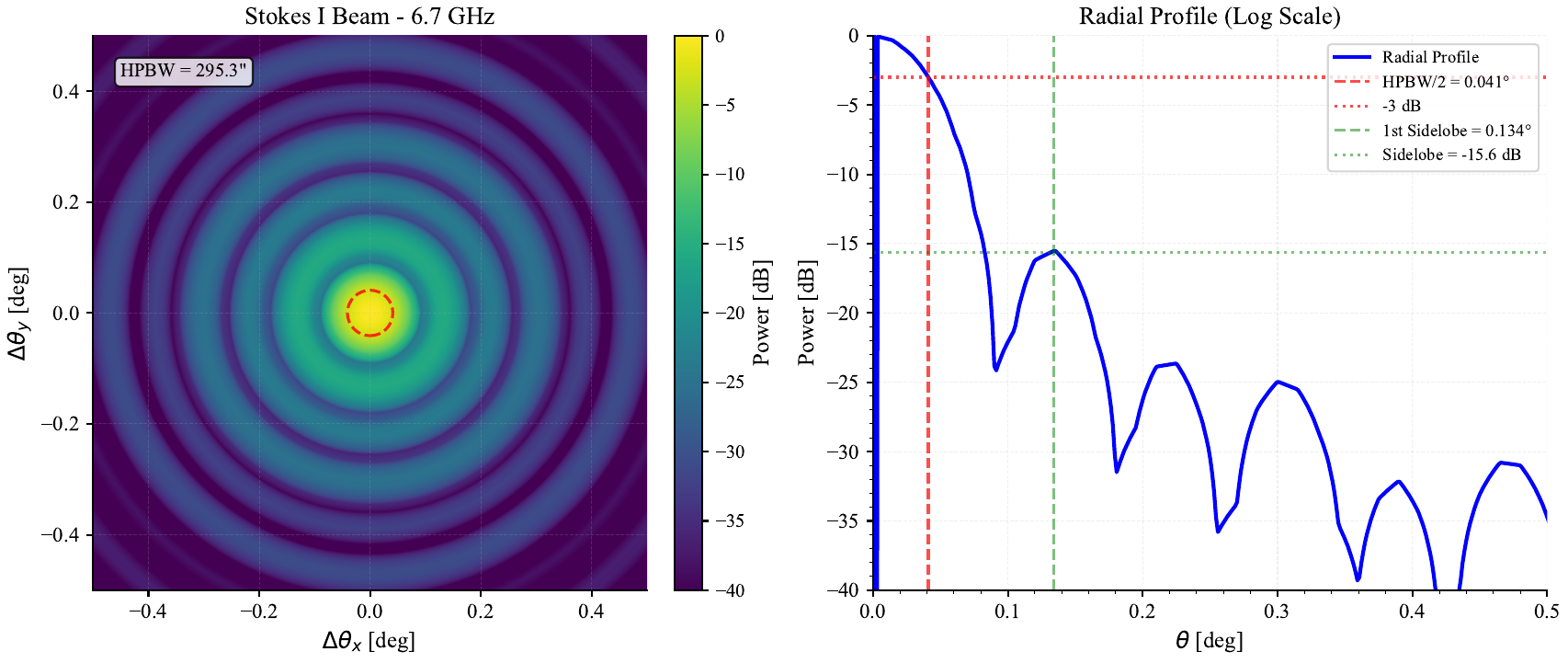}
\caption{}
\label{fig:stokes_I_analysis_67}
\end{subfigure}
\caption{
Stokes $I$ primary beam characteristics derived from the \texttt{GRASP} simulations.
(a) 5.0~GHz: two-dimensional normalised Stokes $I$ beam shown in logarithmic power units (dB), with the dashed circle indicating the half-power beam width (HPBW = 396.4~arcsec). The accompanying azimuthally averaged radial profile highlights the $-3$~dB point and the location and level of the first sidelobe.
(b) 6.7~GHz: corresponding Stokes $I$ beam and radial profile, showing a narrower main lobe (HPBW $= 295.3$~arcsec) and the inward shift of the first sidelobe with increasing frequency. In both cases, the radial profiles provide a quantitative reference for defining the angular scales relevant to subsequent polarimetric analysis.
}
\label{fig:stokes_I_analysis}
\end{figure*}

Figures~\ref{fig:stokes_I_analysis_5} and~\ref{fig:stokes_I_analysis_67} present the Stokes $I$ primary beam responses at $5.0$ and $6.7$~GHz, providing a direct measure of the angular sensitivity of the antenna to unpolarised radiation. In both cases, the beam exhibits a single, well-defined maximum at the nominal boresight and a high degree of azimuthal symmetry across the main lobe. The absence of discernible ellipticity, secondary maxima, or azimuthally localised distortions within the central field of view indicates that the optical response is dominated by a single, stable illumination mode. This behaviour is essential for subsequent polarimetric analysis, as asymmetries in the total-power beam are a primary source of direction-dependent leakage through differential coupling of the orthogonal field components.

The radial beam profiles reveal a smooth and monotonic decline in power from the peak to the half-power point, followed by a sequence of clearly resolved diffraction sidelobes. At $5.0$~GHz, the half-power beam width is measured to be $396.4$~arcsec ($0.110$~deg), while at $6.7$~GHz it reduces to $295.3$~arcsec ($0.082$~deg). This frequency dependence is consistent with the expected inverse scaling of angular resolution for a fixed physical aperture and indicates that the simulated beams operate close to the diffraction limit. The regularity of the main-lobe structure suggests that flux calibration based on point sources is unlikely to be compromised by fine-scale beam features within the half-power radius.

Beyond the main lobe, the first sidelobe is detected at angular offsets of approximately $0.178$~deg at $5.0$~GHz and $0.134$~deg at $6.7$~GHz, with corresponding amplitudes of $-15.4$~dB and $-15.6$~dB relative to the peak. The close agreement in sidelobe level across frequency implies that the effective illumination taper and edge truncation of the aperture remain broadly stable over the band considered. These sidelobe characteristics define the angular scale at which off-axis response becomes non-negligible and therefore delimit the region over which compact-source measurements are primarily governed by the main beam rather than by far-out power redistribution.

By integrating the Stokes $I$ response within a radius of approximately $1.2$ times the half-power beam width, the main-beam efficiency is estimated to be $0.564$ at $5.0$~GHz and $0.556$ at $6.7$~GHz. This indicates that slightly more than half of the total collected power is concentrated within the main beam, with the remainder distributed into sidelobes and the near-out region. From the standpoint of polarimetric performance, this localisation of power is significant: for compact sources, the dominant contribution to both instrumental leakage and cross-polar coupling arises within the main beam, while sidelobe structure increasingly influences measurements of extended emission or sources located away from the pointing centre. The Stokes $I$ beam therefore establishes the angular and energetic framework within which the field-level Jones responses are interpreted in the following analysis.

\begin{figure*}
\begin{minipage}[H]{\linewidth}
\centering
\includegraphics[width=\textwidth]{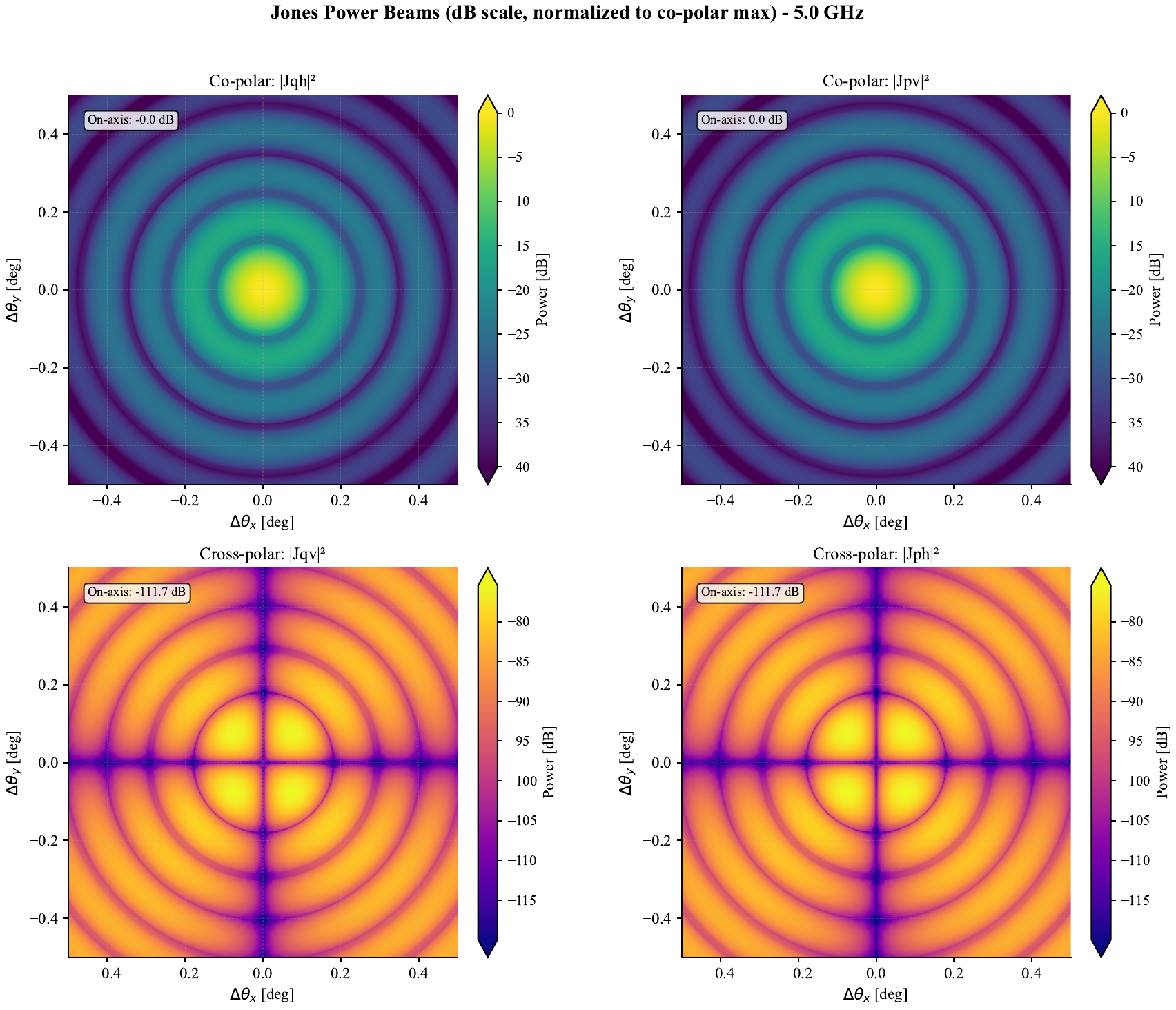} 
\end{minipage}
\caption{
Jones power beams at $5.0$~GHz, displayed on a logarithmic scale and normalised to the peak co-polar response. The upper panels show the co-polar components $|J_{qh}|^2$ and $|J_{pv}|^2$, which exhibit centrally peaked, near-identical beam patterns aligned with the telescope boresight. The lower panels show the cross-polar components $|J_{qv}|^2$ and $|J_{ph}|^2$, which are strongly suppressed at the beam centre and increase with angular offset, forming a symmetric four-lobed structure. On-axis cross-polar power reaches approximately $-112$~dB relative to the co-polar peak, while levels within the half-power beam width rise to the $-70$ to $-80$~dB range. All panels are shown as functions of angular offset from boresight.
}
\label{fig:jones_power_beams}
\end{figure*}

The field-level structure underlying the total-power response is examined through the Jones power beams shown in Fig.~\ref{fig:jones_power_beams}. The co-polar components, $|J_{qh}|^2$ and $|J_{pv}|^2$, exhibit beam patterns that closely mirror the Stokes $I$ response discussed above, with centrally peaked main lobes and concentric diffraction rings. Both components attain their maxima at the nominal boresight and are normalised to $0$~dB on-axis, confirming that the two orthogonal linear polarisations couple symmetrically to the reflector within numerical precision. The near-identical morphology and radial extent of the two co-polar beams indicate that differential gain between the nominally orthogonal feeds is negligible across the main beam, a prerequisite for stable recovery of linear polarisation.

The cross-polar power beams, $|J_{qv}|^2$ and $|J_{ph}|^2$, display a markedly different spatial structure. On-axis cross-polar levels reach approximately $-112$~dB relative to the co-polar peak, reflecting the near-orthogonality of the feed response at the beam centre. This extremely low central value should be interpreted as a numerical lower bound rather than a physically meaningful suppression level, arising from the vanishing of the cross-polar field at exact boresight. Away from the centre, the cross-polar response increases rapidly and adopts a characteristic four-lobed pattern aligned with the principal axes, consistent with symmetry-driven coupling between orthogonal field components in an axisymmetric optical system. At angular offsets comparable to the half-power beam width, the cross-polar power rises to the $-70$ to $-80$~dB range, before continuing to increase towards the sidelobe region.

The spatial segregation between the co-polar main lobe and the cross-polar response has two important consequences. First, it confirms that the dominant contribution to cross-polarisation within the main beam arises from off-axis effects rather than from intrinsic feed leakage at boresight. Second, it demonstrates that any metric based solely on on-axis ratios, such as the raw cross-polarisation ratio, substantially overestimates the polarimetric purity achievable across the beam. Instead, the rapid growth of cross-polar power with angular offset implies that direction-dependent effects must be assessed over a finite region of the beam, particularly within the half-power radius identified from the Stokes $I$ analysis.

The relative symmetry between $|J_{qv}|^2$ and $|J_{ph}|^2$ further indicates that cross-coupling is balanced between the two linear channels, with no evidence for a preferred leakage direction. This balance is reflected in the quadrupolar structure of the cross-polar beams and sets the stage for interpreting instrumental polarisation in terms of combined Jones-element behaviour rather than isolated terms. These field-level characteristics motivate the transition to Mueller-based quantities in the following analysis, where the impact of cross-polar coupling on the recovered Stokes parameters can be assessed directly.

\begin{figure*}
\begin{minipage}[H]{\linewidth}
\centering
\includegraphics[width=\textwidth]{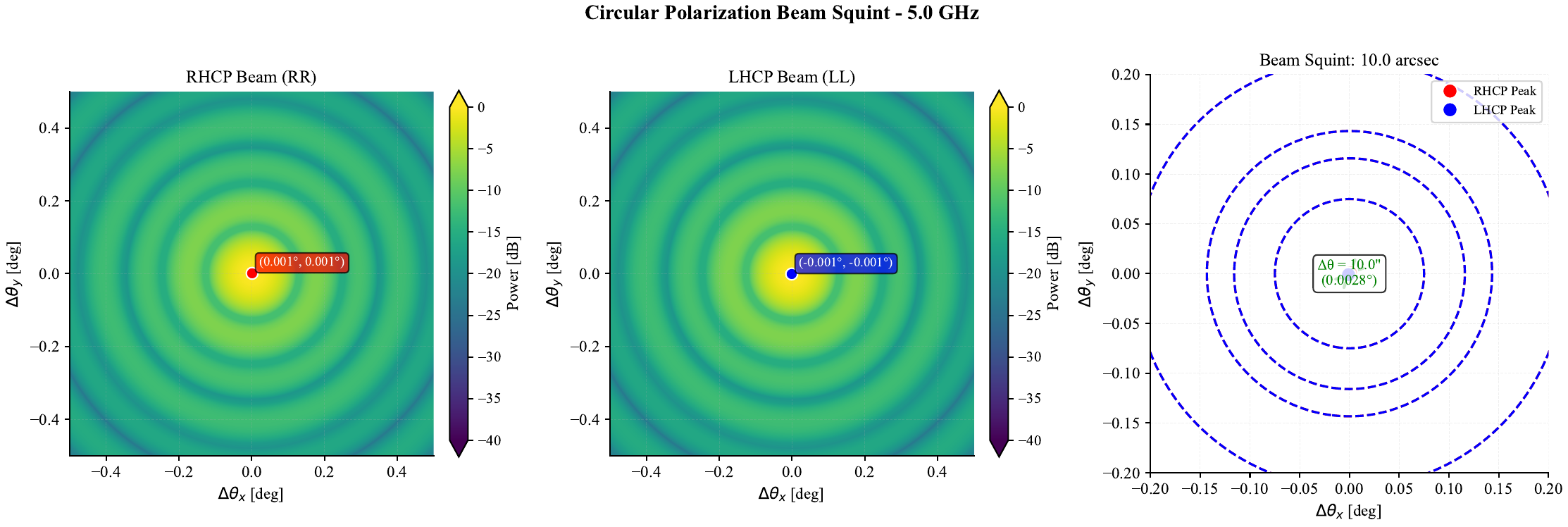} 
\end{minipage}
\caption{
Circular polarisation beam squint at $5.0$~GHz. 
Left and centre panels show the RHCP (RR) and LHCP (LL) power beams, respectively, each normalised to its own peak. 
The right panel highlights the relative displacement between the RHCP and LHCP beam centres, yielding a measured squint of $10.0$~arcsec ($2.8\times10^{-3}$~deg). 
Although small compared to the full beam extent, this offset represents a systematic geometric effect that contributes to off-axis polarimetric leakage and degradation of polarimetric conditioning.
}
\label{fig:beam_squint_5}
\end{figure*}

Having established the angular structure of the Jones fields, Fig.~\ref{fig:beam_squint_5} examines the relative displacement of the right- and left-hand circularly polarised beams. The RHCP and LHCP power peaks are offset from the nominal boresight by $(+0.001^\circ,+0.001^\circ)$ and $(-0.001^\circ,-0.001^\circ)$, respectively, corresponding to an absolute beam squint of $10.0$~arcsec ($2.8\times10^{-3}$~deg). When expressed relative to the measured half-power beam width, this offset represents only a small fraction of the main beam extent. The squint is therefore visually subtle on the full beam maps, yet geometrically significant: even sub-percent displacements of the circular beams introduce systematic phase gradients that couple total intensity into polarisation off axis.

This measured squint provides a direct physical link between the Jones-level structure and the polarimetric metrics that follow. Beam squint shifts the effective pointing of the two circular components relative to one another, so that an unpolarised source observed away from boresight is sampled by unequal portions of the primary beam. This mechanism produces direction-dependent leakage and contributes to the degradation of polarimetric conditioning, particularly toward the half-power radius and beyond.

\begin{figure*}
\begin{minipage}[H]{\linewidth}
\centering
\includegraphics[width=\textwidth]{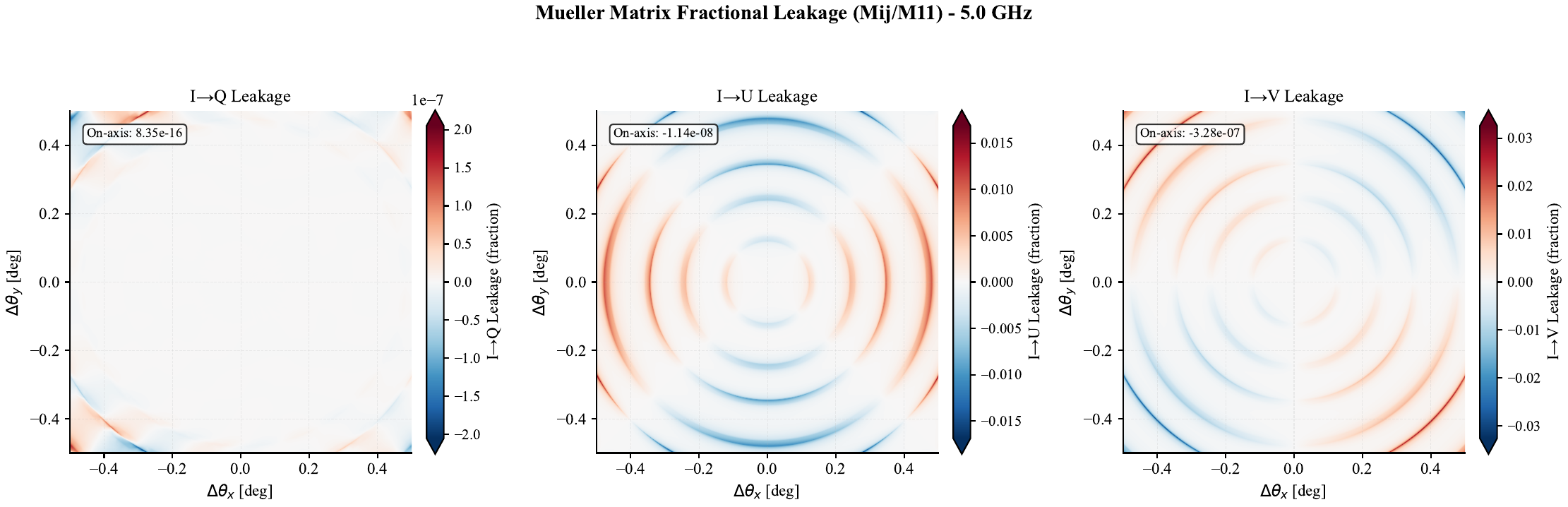} 
\end{minipage}
\caption{
Fractional Mueller matrix leakage terms at $5.0$~GHz, normalised by the Stokes~$I$ response. 
The panels show leakage from total intensity into Stokes $Q$ ($M_{21}/M_{11}$), $U$ ($M_{31}/M_{11}$), and $V$ ($M_{41}/M_{11}$). 
All leakage terms are negligible at boresight but increase with angular offset, exhibiting ring-like structures correlated with the sidelobes of the primary beam. 
The combined linear instrumental polarisation remains below $\sim0.5$~per~cent within the half-power beam width and rises sharply in the sidelobe region.
}
\label{fig:mueller_leakage_5}
\end{figure*}

The observable consequences of these field-level effects are shown in Figure~\ref{fig:mueller_leakage_5}, which presents the fractional Mueller leakage terms normalised by the Stokes~$I$ response. The $I\rightarrow Q$ and $I \rightarrow U$ terms, $M_{21}/M_{11}$ and $M_{31}/M_{11}$, are effectively zero at boresight, with on-axis values below $10^{-7}$ in fractional units. Off axis, both terms exhibit concentric ring-like structures that closely trace the sidelobes of the total-power beam, reaching peak fractional leakages of order $10^{-2}$ to $10^{-3}$ at angular offsets approaching the edge of the plotted field. This behaviour demonstrates that instrumental linear polarisation is not random, but strongly modulated by the underlying beam structure.

The $I \rightarrow V$ term, $M_{41}/M_{11}$, shows a similar radial dependence but with a distinct sign structure, reflecting the additional role of relative phase between the orthogonal field components. Although the absolute magnitude of circular leakage remains small, its systematic spatial variation indicates that beam squint and cross-polar coupling jointly influence the recovery of Stokes~$V$, particularly for off-axis sources. The combined linear instrumental polarisation, defined as $\sqrt{M_{21}^2+M_{31}^2}/M_{11}$, remains below $0.5$~per~cent within the half-power beam width but rises sharply in the sidelobes, underscoring the importance of beam modelling for wide-field or extended-source observation

\begin{figure*}
\begin{minipage}[H]{\linewidth}
\centering
\includegraphics[width=\textwidth]{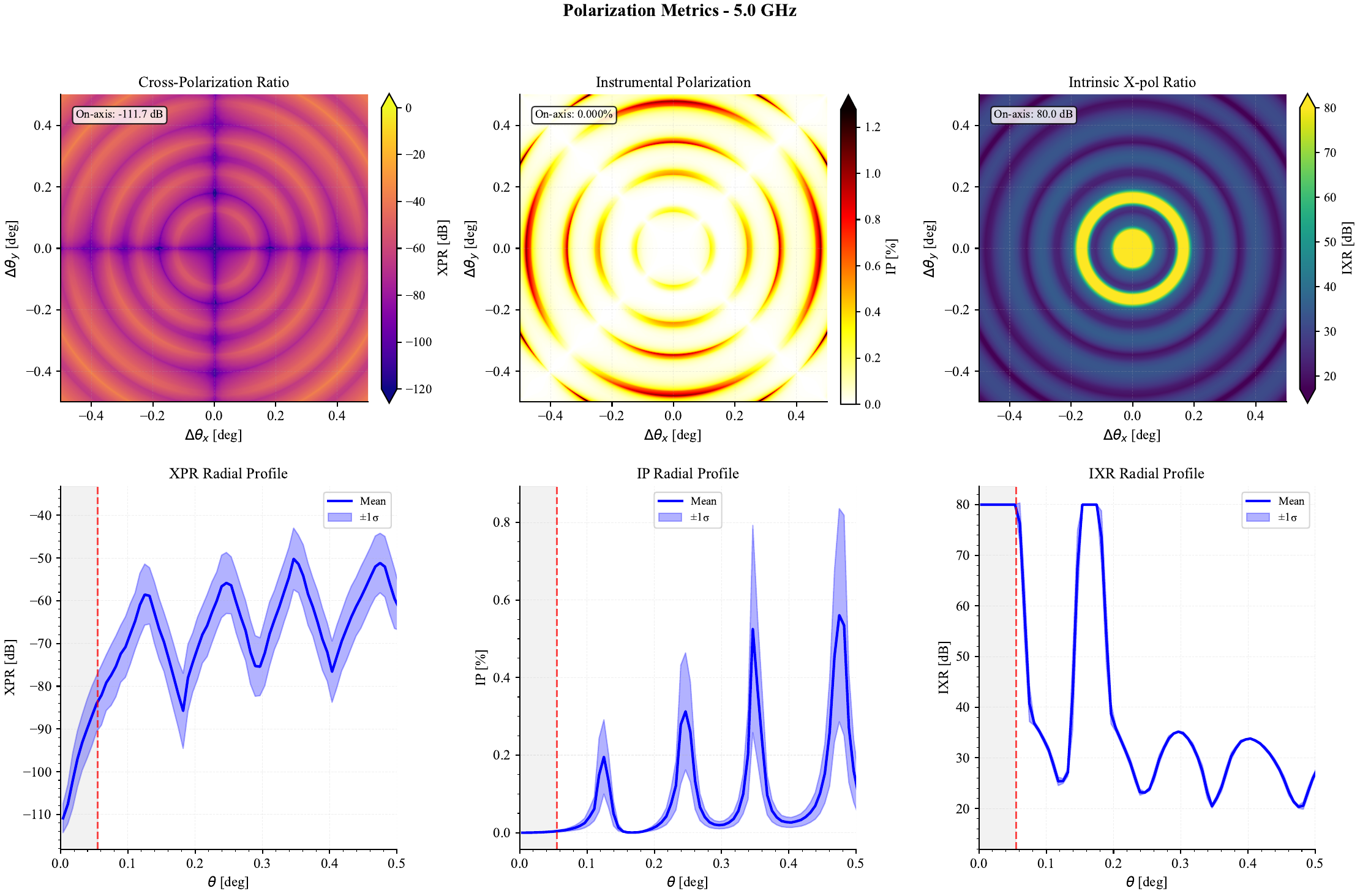} 
\end{minipage}
\caption{
Derived polarimetric performance metrics at $5.0$~GHz. 
Top row: two-dimensional maps of the XPR, IP, and IXR. 
Bottom row: corresponding radial profiles showing the mean behaviour and dispersion as a function of angular offset. 
XPR and IXR are highest at boresight and degrade rapidly with offset, while IP increases smoothly across the main beam and becomes dominated by sidelobe structure beyond the half-power radius.
}
\label{fig:pol_metrics_5}
\end{figure*}

Figure~\ref{fig:pol_metrics_5} summarises the cumulative impact of these effects through derived polarimetric performance metrics. The cross-polarisation ratio (XPR) map reflects the strong suppression of cross-polar power on axis, with values below $-110$~dB at boresight, followed by a rapid degradation with increasing offset. The corresponding radial profile shows that XPR decreases by more than $40$~dB between the beam centre and the half-power radius, highlighting the limited angular region over which extremely high polarisation purity is maintained.
The instrumental polarisation (IP) metric provides a more directly observable measure. Its radial profile increases smoothly from values consistent with numerical noise at boresight to mean levels of $\sim0.3$–$0.5$~per~cent near the half-power radius, with substantial scatter introduced by sidelobe structure. This behaviour confirms that instrumental polarisation is intrinsically direction-dependent and cannot be characterised by a single global number.
Finally, the IXR quantifies the conditioning of the polarimetric response. On-axis, IXR reaches $\sim80$~dB, indicating excellent numerical stability for polarisation inversion. Away from the beam centre, IXR declines sharply, with minima of $\sim25$–$30$~dB near the first sidelobe. These regions correspond closely to where cross-polar power, beam squint, and Mueller leakage are all enhanced, demonstrating that the degradation of polarimetric fidelity is driven by a combination of geometric and field-level effects rather than by any single mechanism in isolation.

\begin{table*}
\centering
\caption{Fractional instrumental polarisation as a function of angular
offset at $5.0$ and $6.7$~GHz.}
\label{tab:ip_offset}
\begin{tabular}{lcccccc}
\hline
Frequency & Angular region & Radius (arcsec) & Max IP & Mean IP & Std IP & Min IP \\
(GHz) & & & (\%) & (\%) & (\%) & (\%) \\
\hline
5.0 & HPBW/4  & 99.1  & 0.00124 & 0.00044 & 0.00034 & $1\times10^{-6}$ \\
5.0 & HPBW/2  & 198.2 & 0.00576 & 0.00182 & 0.00151 & $1\times10^{-6}$ \\
5.0 & HPBW    & 396.4 & 0.08689 & 0.01602 & 0.01891 & $<10^{-6}$ \\
5.0 & 0.20$^\circ$ & 720.0 & 0.36874 & 0.03247 & 0.06274 & $<10^{-6}$ \\
\hline
6.7 & HPBW/4  & 73.8  & 0.00025 & 0.00010 & 0.00007 & $<10^{-6}$ \\
6.7 & HPBW/2  & 147.7 & 0.00142 & 0.00044 & 0.00037 & $<10^{-6}$ \\
6.7 & HPBW    & 295.3 & 0.03037 & 0.00521 & 0.00664 & $<10^{-6}$ \\
6.7 & 0.20$^\circ$ & 720.0 & 0.47750 & 0.05406 & 0.08031 & $<10^{-6}$ \\
\hline
\end{tabular}
\end{table*}

\begin{table}
\centering
\caption{IXR statistics evaluated over distinct beam regions at $5.0$ and $6.7$~GHz. 
Values are given in dB and summarise the spatial variation of polarimetric conditioning.}
\label{tab:ixr_summary}
\begin{tabular}{cccccc}
\hline
Frequency & Region & Min & Mean & Max & Std \\
(GHz) &  & (dB) & (dB) & (dB) & (dB) \\
\hline
5.0 & On-axis       & 80.0 & 80.0 & 80.0 & 0.0 \\
5.0 & HPBW/4        & 80.0 & 80.0 & 80.0 & 0.0 \\
5.0 & HPBW/2        & 80.0 & 80.0 & 80.0 & 0.0 \\
5.0 & HPBW          & 28.9 & 51.7 & 80.0 & 21.0 \\
5.0 & First sidelobe& 22.6 & 39.6 & 80.0 & 18.9 \\
\hline
6.7 & On-axis       & 80.0 & 80.0 & 80.0 & 0.0 \\
6.7 & HPBW/4        & 80.0 & 80.0 & 80.0 & 0.0 \\
6.7 & HPBW/2        & 80.0 & 80.0 & 80.0 & 0.0 \\
6.7 & HPBW          & 32.0 & 65.7 & 80.0 & 18.4 \\
6.7 & First sidelobe& 23.1 & 33.8 & 79.9 & 10.0 \\
\hline
\end{tabular}
\end{table}

The quantitative impact of direction-dependent instrumental
polarisation is summarised in Table~\ref{tab:ip_offset}. The statistics
show that the leakage response is strongly dependent on angular
position, with substantially greater azimuthal variation toward the
outer beam. This behaviour indicates that instrumental polarisation
cannot be represented adequately by a single radial or on-axis value.

The corresponding regional IXR statistics are given in
Table~\ref{tab:ixr_summary}. The response is highly conditioned in the
central beam but becomes increasingly variable toward the half-power
boundary and beyond. The higher mean IXR at $6.7$~GHz indicates that
the polarimetric response remains better conditioned over the central
beam at the higher observing frequency.

A consolidated overview of the principal beam and polarimetric
properties is provided in Table~\ref{tab:beam_metrics}. The computed
half-power beam widths are $396.4$~arcsec ($0.1101^\circ$) at
$5.0$~GHz and $295.3$~arcsec ($0.0820^\circ$), respectively, while the
main-beam efficiencies are $0.564$ and $0.556$. The circular-polarisation
beam centres are displaced by $9.96$~arcsec at $5.0$~GHz, corresponding
to approximately $2.5$~per~cent of the HPBW. At $6.7$~GHz, no
resolvable displacement is obtained at the $7.03$~arcsec~pixel$^{-1}$
sampling of the tangent-plane map, giving an upper limit of
$<7.03$~arcsec. These quantities provide the principal beam-scale
parameters used in interpreting the direction-dependent
polarimetric response.

As an internal consistency check, the total-intensity beam
characteristics were compared with the earlier electromagnetic
characterisation of the same GRAO 32-m shaped dual-reflector system by
\citet{venter2018electromagnetic}. Their \texttt{GRASP} analysis
reported half-power beam widths of approximately $0.11^\circ$ and
$0.09^\circ$ at $5.0$ and $6.7$~GHz, respectively, with first
sidelobe levels of $-15.21$ and $-15.15$~dB. The present model gives
corresponding beam widths of $0.1101^\circ$ and $0.0820^\circ$, and
first sidelobe levels of approximately $-15.4$ and $-15.6$~dB. The
close correspondence provides an independent consistency check on the
underlying electromagnetic description of the telescope. The earlier
study also identified asymmetric illumination and enhanced sidelobe
structure associated with the shaped reflector and slanted
beam-waveguide \citep{venter2018electromagnetic}, providing physical
context for the direction-dependent structure resolved in the present
polarimetric analysis.

\begin{table}
\centering
\caption{Summary of computed beam and polarimetric performance metrics
at $5.0$ and $6.7$~GHz.}
\label{tab:beam_metrics}
\begin{tabular}{lcc}
\hline
Metric & 5.0 GHz & 6.7 GHz \\
\hline
HPBW (deg)                    & 0.1101 & 0.0820 \\
HPBW (arcsec)                 & 396.4  & 295.3  \\
Main-beam efficiency          & 0.564  & 0.556  \\
Beam squint (arcsec)          & 9.96   & $<7.03$ \\
Beam squint / HPBW (\%)       & 2.51   & $<2.38$ \\
On-axis XPR (dB)              & $-111.7$ & $-109.0$ \\
On-axis IP (\%)               & $<10^{-6}$ & $<10^{-6}$ \\
On-axis IXR (dB)              & 80.0   & 80.0   \\
Peak Stokes $I$               & 1.0    & 1.0    \\
Dynamic range (dB)            & 47.5   & 48.9   \\
\hline
\end{tabular}
\end{table}

\section{Discussion}
\label{sec:discussion}

The principal implication of the present analysis is that the
polarimetric response of the GRAO 32-m telescope is intrinsically
direction dependent. The electromagnetic model therefore indicates
that polarimetric performance cannot be adequately represented by a
single telescope-wide leakage coefficient or by an on-axis
characterisation alone. Instead, the response must be considered as a
spatially varying property of the primary beam, with the Jones and
Mueller representations providing complementary descriptions of the
same direction-dependent instrumental response. This distinction is
particularly relevant to observations in which the source occupies a
substantial fraction of the beam or is sampled at different positions
during mapping.

The physical origin of this behaviour is consistent with the
electromagnetic response expected from a reflector-based polarimetric
system. Close to boresight, the optical and feed configuration
preserves a high degree of orthogonality between the two polarisation
responses. With increasing angular displacement, however, the
direction-dependent amplitudes and phases of the complex field
responses change, producing increasing coupling between the
polarisation states. The resulting instrumental polarisation is
therefore inherently spatial rather than a fixed property of the
receiver. Similar direction-dependent behaviour has been reported for
other single-dish radio telescopes
\citep{ng2005polarization, Heiles2001}, supporting the interpretation
that the off-axis structure obtained for GRAO is a consequence of the
antenna's electromagnetic response rather than an anomalous feature of
the present analysis.

This spatial character has a direct consequence for calibration. An
on-axis calibration is appropriate for characterising the response at
the pointing centre, but it cannot in general remove leakage produced
when emission is observed through other parts of the beam. The
limitation becomes especially important for extended emission,
raster-scan observations, and other measurements in which different
regions of the sky are sampled through different beam positions. A
direction-dependent beam model can instead be incorporated into the
measurement equation so that the instrumental response is evaluated at
the relevant source position. Observational studies of full-Stokes
single-dish systems have demonstrated the importance of accounting for
beam-dependent instrumental polarisation in achieving accurate
polarimetric measurements \citep{Myserlis2018}. The Jones and Mueller
descriptions developed here consequently provide a suitable
electromagnetic basis for such a calibration framework.

Beam squint represents a related but distinct direction-dependent
effect. Differential displacement of the two circular-polarisation
beam responses means that the receptors can sample different parts of
a source even when their individual beam shapes are otherwise well
characterised. For compact sources close to the pointing centre, this
effect can be relatively limited, whereas for extended emission or
scanning observations it can couple spatial structure in the source to
the measured polarisation. Spatial gradients can consequently be
mapped into apparent polarisation structure, including spurious
circular-polarisation signatures. Such effects are well established in
single-dish polarimetry \citep{carretti2005polarized,
manchester2001parkes}. The frequency dependence found for GRAO
therefore suggests that beam squint should be treated as part of the
direction-dependent instrumental response rather than solely as a
geometrical beam parameter.

The conditioning of the polarimetric response provides a complementary
way of assessing this behaviour. Unlike instrumental polarisation,
which describes the magnitude of unwanted coupling in the observable
Stokes response, IXR characterises the stability of the inversion of
the polarisation response. The high intrinsic conditioning near
boresight indicates that the electromagnetic system provides a robust
basis for recovering incident polarisation when the source is centred
on the beam. The progressive reduction in conditioning towards larger
offsets is more important from a calibration perspective because
polarimetric inversion becomes increasingly sensitive to uncertainties
in the instrumental response and to measurement noise. This distinction
is important when interpreting telescope performance: a favourable
on-axis polarimetric response does not imply uniform polarimetric
conditioning throughout the field of view. The simulations of
\citet{foster2015intrinsic} similarly demonstrate that finite
polarimetric purity can propagate into errors in recovered astronomical
signals, although the practical significance of a given IXR value is
application dependent.

The broader literature provides useful context for the GRAO results,
but does not support a simple numerical ranking of polarimetric
performance among different radio telescopes. Published studies often
characterise different components or stages of the instrumental chain
and use different metrics. For example, Effelsberg measurements have
demonstrated sub-percent polarimetric accuracy after correction of
instrumental effects across the beam \citep{Myserlis2018}, while
Parkes measurements have reported receptor-level IXR values over a
broad observing band \citep{Hobbs2020}. These results demonstrate the
level of polarimetric performance achievable in established
single-dish systems, but they are not directly equivalent to the
intrinsic, direction-dependent GRAO electromagnetic quantities derived
here. Likewise, studies of Arecibo, the DRAO Synthesis Telescope, and
MeerKAT demonstrate that instrumental polarisation and polarimetric
response can vary substantially across the beam
\citep{deVilliers2022, ng2005polarization, Heiles2001}. A common
direction-dependent IXR benchmark across facilities is not yet
established, making metric-, frequency-, and calibration-matched
comparisons preferable to simple telescope-to-telescope ranking.

An important consideration is therefore the extent to which the
simulated electromagnetic response represents the realised telescope.
The comparison by \citet{Asad2021} between a \texttt{GRASP} model and
full-polarisation astro-holographic measurements of MeerKAT provides a
useful empirical example. Their substantially closer agreement for the
diagonal Jones terms than for the weaker off-diagonal terms illustrates
that cross-polarisation quantities are intrinsically more demanding to
validate. Small absolute differences in weak cross-polar responses can
produce comparatively large relative differences, while measurement
noise and uncertainty can also become important. Consequently, good
agreement in the dominant co-polar beam does not by itself establish
the accuracy of the corresponding cross-polarisation or polarimetric
purity prediction.

This comparison does not justify applying a universal empirical
degradation factor to electromagnetic predictions. The difference
between a simulated and realised response depends on the particular
metric, frequency, angular region, measurement technique, and state of
the instrument. The appropriate interpretation of the present GRAO
results is therefore as an intrinsic electromagnetic reference for the
adopted telescope configuration. The realised observational response
may additionally contain contributions from receiver gain and phase
variations and other signal-chain effects, as well as pointing,
alignment, atmospheric, temporal, and physical-model uncertainties.
These effects are not represented by the present electromagnetic beam
calculation and should be distinguished from the intrinsic beam
response when assessing absolute on-sky polarimetric accuracy.

This distinction is particularly relevant to the current development
stage of the GRAO 32-m telescope. The absence of on-sky measurements
in the present work defines the scope of the study rather than
preventing an electromagnetic characterisation of the telescope.
The simulations establish the direction-dependent response expected
from the adopted optical and electromagnetic configuration before the
complete realised instrumental system is available for routine
observations. During commissioning, targeted polarimetric observations
and, where feasible, direct beam measurements can test the predicted
spatial response and determine the additional contributions introduced
by the realised receiver, telescope configuration, and observing
conditions. Such measurements will provide the appropriate bridge
between the intrinsic electromagnetic model developed here and the
eventual end-to-end calibration of the telescope.

These findings establish a physically motivated framework
for understanding GRAO polarimetry in which beam structure,
polarisation coupling, beam squint, and polarimetric conditioning are
treated as components of a single direction-dependent instrumental
response. The scientific implication is that the polarimetric
capability of the telescope should be evaluated over the relevant
field of view rather than by a single on-axis performance measure. The
present electromagnetic model consequently provides the baseline
required for subsequent commissioning validation and for the
development of direction-dependent polarimetric calibration of the
GRAO 32-m telescope.

\section{Conclusion}
\label{sec:conclusion}

This study establishes a direction-dependent polarimetric beam
characterisation of the GRAO 32-m telescope based on full-wave
electromagnetic simulations and a unified Jones--Mueller framework.
The principal contribution is the preservation of the complex
polarisation response from the electromagnetic field level through to
Stokes-space observables, providing a consistent basis for describing
the telescope response across the primary beam rather than reducing
polarimetric performance to a single on-axis leakage parameter.

The analysis establishes that the polarimetric response of the GRAO
32-m is intrinsically direction-dependent. Consequently, polarimetric
fidelity cannot be fully represented by a single scalar beam or
calibration coefficient: the response varies with position within and
beyond the main beam and must therefore be treated as part of the
spatially varying instrumental response of the telescope. The resulting
Jones and Mueller representations provide a common framework in which
beam shape, polarisation coupling, and matrix conditioning can be
considered together.

The main outcome of this work is therefore not only a set of beam
characteristics for the two frequencies considered, but an
electromagnetic reference model against which the polarimetric
behaviour of the GRAO 32-m can be assessed. This provides the necessary
foundation for incorporating direction-dependent instrumental effects
into future polarimetric calibration and data analysis, particularly
for observations in which the source distribution extends
significantly away from the pointing centre.

The model should be regarded as the intrinsic electromagnetic baseline
of the adopted telescope configuration rather than a complete
description of realised on-sky performance. Its principal value is
that it establishes the expected direction-dependent response before
additional observational and instrumental effects are introduced.
Future commissioning measurements can therefore use this model as a
reference for validating the telescope response and for separating
electromagnetic beam behaviour from effects introduced by the receiver,
observing system, and operational state of the telescope. In this
sense, the present work provides a foundation for developing and
validating the polarimetric calibration capability required to fully
exploit the GRAO 32-m telescope for precision radio astronomical
observations.

% \section*{Acknowledgements}

% The authors thank the anonymous referees for their careful reading of the manuscript and for the constructive comments that helped improve its clarity and presentation. T.A.-N is grateful to the research initiative supported by the France–Ghana bilateral program under the Fonds d'Expertise et de Formation – Initiative Africa scheme, which is coordinated by the French Ministry for Europe and Foreign Affairs and implemented through the French Embassy in Ghana. This programme supports collaboration in higher education, research, and innovation between institutions in Ghana and France, and has provided essential funding for this joint academic and scientific project. T.A.-N also gratefully acknowledges the support of the Ghana Space Science and Technology Institute (GSSTI) and access to its High-Performance Computing (HPC) facilities, which were indispensable for carrying out the electromagnetic simulations and data processing presented in this work. The authors further acknowledge the SARAO African VLBI Network (AVN) engineering teams, including the Mechanical and Structural, Control and Monitoring, Timing and Frequency Reference, and Electrical groups, for their continued technical support and dedicated maintenance of the GRAO infrastructure, which underpins the observational context of this study.

\section*{Conflicts of Interest}
The authors declare that they have no competing financial interests or personal relationships that could have appeared to influence the work reported in this paper.

\section*{Acknowledgements}

The authors thank the anonymous referees for their careful reading of the manuscript and for the constructive comments that helped improve its clarity and presentation. T.A.-N is grateful to the Ghana Space Science and Technology Institute (GSSTI) for access to its High-Performance Computing (HPC) facilities, which were indispensable for carrying out the electromagnetic simulations and data processing presented in this work. The authors further acknowledge the SARAO African VLBI Network (AVN) engineering teams, including the Mechanical and Structural, Control and Monitoring, Timing and Frequency Reference, and Electrical groups, for their continued technical support and dedicated maintenance of the GRAO infrastructure, which underpins the observational context of this study.

\section*{Funding}

This research was carried out within the collaborative framework established by the Memorandum of Understanding between the Ghana Atomic Energy Commission (GAEC), acting through the Ghana Space Science and Technology Institute (GSSTI), and the Regents of the University of California on behalf of its Santa Cruz campus (UCSC). Financial support for this work was provided by the GRAO–UCSC Astronomy Development Project and the France–Ghana bilateral program under the Fonds d'Expertise et de Formation – Initiative Africa, coordinated by the French Ministry for Europe and Foreign Affairs and implemented through the French Embassy in Ghana. Institutional resources and computational facilities were provided by the GSSTI.

%%%%%%%%%%%%%%%%%%%%%%%%%%%%%%%%%%%%%%%%%%%%%%%%%%
\section*{Data Availability}

The data underlying this article are derived from electromagnetic simulations of the GRAO 32-m telescope conducted using \texttt{GRASP}.  
These data, together with the associated analysis scripts used to generate the beam patterns and filtering results, will be made available to qualified researchers upon reasonable request to the corresponding author.  
No proprietary or confidential information is included.

% \section*{CRediT authorship contribution statement}

% \textbf{Theophilus Ansah-Narh:} Conceptualization, methodology, software, formal analysis, and writing -- original draft; led the electromagnetic beam modelling, Jones and Mueller matrix formulation, numerical pipeline development, and polarimetric performance analysis. \\

% \textbf{Nia Imara:} Supervision, investigation, and writing -- review and editing; provided astrophysical context and critical interpretation of the polarimetric results, with emphasis on implications for single-dish and VLBI observations. \\

% \textbf{Benedicta Woodea:} Data curation and validation; contributed to the verification of simulation outputs, beam consistency checks, and supporting analysis of polarimetric diagnostics. \\

% \textbf{Joseph Bremang Tandoh:} Software and formal analysis; contributed to the numerical implementation of the data processing pipeline and assisted with result validation and visualisation. \\

% \textbf{Emmanuel Proven-Adzri:} Investigation and resources; supported the electromagnetic modelling framework and provided technical input related to the GRAO system configuration and simulation environment.\\

% \textbf{Diana Klutse:} Supported the electromagnetic modelling framework.\\

% \textbf{Emmanuel Proven-Adzri:} Investigation and resources; supported the electromagnetic modelling framework and provided technical input related to the GRAO system configuration and simulation environment.



\bibliographystyle{rasti}
\bibliography{example} % if your bibtex file is called example.bib

%\begin{thebibliography}{99}
%\end{thebibliography}

%%%%%%%%%%%%%%%%%%%%%%%%%%%%%%%%%%%%%%%%%%%%%%%%%%

%%%%%%%%%%%%%%%%% APPENDICES %%%%%%%%%%%%%%%%%%%%%

% \appendix

% \section{Some extra material}

% If you want to present additional material which would interrupt the flow of the main paper,
% it can be placed in an Appendix which appears after the list of references.

% %%%%%%%%%%%%%%%%%%%%%%%%%%%%%%%%%%%%%%%%%%%%%%%%%%

% Don't change these lines
\bsp	% typesetting comment
\label{lastpage}
\end{document}